\documentclass[a4paper,11pt]{article}
\pdfoutput=1 

\usepackage{jheppub} 
\makeatletter
\DeclareRobustCommand*{\bfseries}{%
  \not@math@alphabet\bfseries\mathbf
  \fontseries\bfdefault\selectfont
  \boldmath
}
\makeatother

\usepackage{calc}
\usepackage{rotating}
\usepackage[english]{babel}
\usepackage{graphicx}
\usepackage{subfig}
\usepackage{float}
\usepackage{amsmath}
\usepackage{amssymb}
\usepackage{amsthm}
\usepackage{latexsym}
\usepackage{dcolumn}
\usepackage{hyperref}
\newcommand{\newc}{\newcommand*}

\long\def\begincomment#1\endcomment{%
        \begingroup\sf\baselineskip12pt#1\endgroup}

\newc{\etal}{\textrm{et al.}} 
\newc{\eg}{\textrm{e.g.}} 
\newc{\ie}{\textrm{i.e.}}
\newc{\etc}{\textrm{etc}}
\newc\vs{\textrm{vs.}}
\newc{\cl}{\rm {C.L.}}

\newc{\ev}{\ensuremath{\,\mathrm{eV}}}
\newc{\kev}{\ensuremath{\,\mathrm{keV}}}
\newc{\mev}{\ensuremath{\,\mathrm{MeV}}}
\newc{\gev}{\ensuremath{\,\mathrm{GeV}}}
\newc{\tev}{\ensuremath{\,\mathrm{TeV}}}
\newc{\MeV}{\mev} 
\newc{\TeV}{\tev}
\newc{\invpb}{\ensuremath{/\text{pb}}}
\newc{\invfb}{\ensuremath{\,\text{fb}^{-1}}}
\newc\nb{\ensuremath{\,\mathrm{nb}}} \newc\pb{\ensuremath{\,\mathrm{pb}}} \newc\fb{\ensuremath{\,\mathrm{fb}}}
\newc\pc{\ensuremath{\,\mathrm{pc}}}
\newc\kpc{\ensuremath{\,\mathrm{kpc}}}
\newc\mpc{\ensuremath{\,\mathrm{Mpc}}}
\newc\ps{\ensuremath{\,\mathrm{ps}}} 
\newc\cmeter{\ensuremath{\,\mathrm{cm}}} 
\newc\meter{\ensuremath{\,\mathrm{m}}} 
\newc\kmeter{\ensuremath{\,\mathrm{km}}}
\newc\second{\ensuremath{\,\mathrm{s}}}
\newc\msecond{\ensuremath{\,\mathrm{ms}}}
\newc\nsecond{\ensuremath{\,\mathrm{ns}}}
\newc\psecond{\ensuremath{\,\mathrm{ps}}}

\newc{\chisqmin}{\ensuremath{\chi^2_{\mathrm{min}}}}
\newc{\Delchisq}{\ensuremath{\Delta\chi^2}}
\newc{\chisq}{\ensuremath{\chi^2}}
\newc{\like}{\ensuremath{\mathcal{L}}}

\newc\lsim{\ensuremath{\mathrel{\rlap{\lower4pt\hbox{\hskip1pt$\sim$}}\raise1pt\hbox{$<$}}}}
\newc\gsim{\ensuremath{\mathrel{\rlap{\lower4pt\hbox{\hskip1pt$\sim$}}\raise1pt\hbox{$>$}}}}
\newc{\VEV}[1]{\ensuremath{\langle #1 \rangle}}
\newc{\dl}{\ensuremath{\stackrel{\leftarrow}{D}}}
\newc{\dr}{\ensuremath{\stackrel{\rightarrow}{D}}}

\newc{\bcenter}{\begin{center}}    \newc{\ecenter}{\end{center}}
\newc{\bfl}{\begin{flushleft}}    \newc{\efl}{\end{flushleft}}
\newc{\bfr}{\begin{flushright}}    \newc{\efr}{\end{flushright}}

\newc{\bi}{\begin{itemize}}
\newc{\ei}{\end{itemize}}
\newc{\bed}{\begin{description}}
\newc{\eed}{\end{description}}
\newc{\ben}{\begin{enumerate}}
\newc{\een}{\end{enumerate}}

\newc{\be}{\begin{equation}}
\newc{\ee}{\end{equation}}
\newc{\bea}{\begin{eqnarray}}
\newc{\eea}{\end{eqnarray}}
\newc{\ra}{\rightarrow}

\newc{\alphas}{\ensuremath{\alpha_s}}
\newc{\alphatwo}{\ensuremath{\alpha_2}}
\newc{\alphaone}{\ensuremath{\alpha_1}}
\newc{\alphai}[1]{\ensuremath{\alpha_{#1}}}
\newc{\alphaem}{\ensuremath{\alpha_{\mathrm{em}}}}
\newc{\alphaeff}{\ensuremath{\alpha_{\mathrm{eff}}}}
\newc{\sineff}{\ensuremath{\sin \theta_{\mathrm{eff}}}}
\newc{\sinsqeff}{\ensuremath{\sin^2 \theta_{\mathrm{eff}}}}
\newc{\dalphahad}{\ensuremath{\Delta \alpha_{\mathrm{had}}}}
\newc{\yt}{\ensuremath{h_t}} \newc{\yb}{\ensuremath{h_b}} \newc{\ytau}{\ensuremath{h_{\tau}}}
\newc\mz{\ensuremath{M_Z}} 
\newc\mw{\ensuremath{m_W}}
\newc\mZ{\mz}        \newc\mW{\mw}
\newc\mhsm{\ensuremath{ m_{H_{\mathrm{SM}}}}}
\newc{\mtop}{\ensuremath{ m_t}}               \newc{\mtpole}{\ensuremath{ M_t}}
\newc{\mbottom}{\ensuremath{ m_b}} 
\newc{\mtau}{\ensuremath{ m_{\tau}}}
\newc{\mt}{\mtpole}
\newc{\mb}{\mbottom} 
\newc{\rgg}{\ensuremath{R_{h}(\gamma\gamma)}}
\newc{\rzz}{\ensuremath{R_{h}(ZZ)}}
\newc{\rtwogg}{\ensuremath{R_{h_2}(\gamma\gamma)}}
\newc{\rtwozz}{\ensuremath{R_{h_2}(ZZ)}}
\newc{\ronegg}{\ensuremath{R_{h_1}(\gamma\gamma)}}
\newc{\ronezz}{\ensuremath{R_{h_1}(ZZ)}}
\newc{\rsiggg}{\ensuremath{R_{h_\textrm{sig}}(\gamma\gamma)}}
\newc{\rsigzz}{\ensuremath{R_{h_\textrm{sig}}(ZZ)}}
\newc{\llbar}{\ensuremath{\ell\bar{\ell}}}
\newc{\tauptaum}{\ensuremath{ \tau^+\tau^-}}
\newc{\qqbar}{\ensuremath{ q\bar{q}}} \newc{\ppbar}{\ensuremath{ p\bar{p}}}
\newc{\bbbar}{\ensuremath{ b\bar{b}}} \newc{\ttbar}{\ensuremath{ t\bar{t}}}
\newc{\ffbar}{\ensuremath{ f\bar{f}}} \newc{\tautaubar}{\ensuremath{ \tau\bar{\tau}}}
\newc{\mchi}{\ensuremath{m_{\chi}}}
\newc{\squark}{\ensuremath{\tilde{q}}}
\newc{\slepton}{\ensuremath{\tilde{l}}}
\newc{\gluino}{\ensuremath{\tilde{g}}} 
\newc{\mgluino}{\ensuremath{{m_{\gluino}}}}
\newc{\tone}{\ensuremath{{\tilde{t}_1}}}

\newc{\sthw}{\ensuremath{ \sin\theta_W}}              \newc{\cthw}{\ensuremath{\cos\theta_W}}
\newc{\tanthw}{\ensuremath{ \tan\theta_W}}              \newc{\cotthw}{\ensuremath{\cot\theta_W}}
\newc{\ssqthw}{\ensuremath{\sin^2 \theta_W}}
\newc{\msbar}{\ensuremath{\overline{MS}}} \newc{\drbar}{\ensuremath{\overline{DR}}}
\newc{\mtmtsmmsbar}{\ensuremath{ m_t(m_t)^{\msbar}_{{\mathrm{SM}}}}}
\newc{\mtmtsmdrbar}{\ensuremath{ m_t(m_t)^{\drbar}_{{\mathrm{SM}}}}}
\newc{\mtmtmssmdrbar}{\ensuremath{ m_t(m_t)^{\drbar}_{{\mathrm{SUSY}}}}}
\newc{\mbmbmsbar}{\ensuremath{ m_b(m_b)^{\msbar} }}
\newc{\mbmbsmmsbar}{\ensuremath{ m_b(m_b)^{\msbar}_{{\mathrm{SM}}}}}
\newc{\mbmzsmmsbar}{\ensuremath{ m_b(\mz)^{\msbar}_{{\mathrm{SM}}}}}
\newc{\mbmzsmdrbar}{\ensuremath{ m_b(\mz)^{\drbar}_{{\mathrm{SM}}}}}
\newc{\mbmzmssmdrbar}{\ensuremath{ m_b(\mz)^{\drbar}_{{\mathrm{SUSY}}}}}
\newc{\mtaumzsmmsbar}{\ensuremath{ m_{\tau}(\mz)^{\msbar}_{{\mathrm{SM}}}}}
\newc{\mtaumzsmdrbar}{\ensuremath{ m_{\tau}(\mz)^{\drbar}_{{\mathrm{SM}}}}}
\newc{\mtaumzmssmdrbar}{\ensuremath{ m_{\tau}(\mz)^{\drbar}_{{\mathrm{SUSY}}}}}
\newc{\alphasmzms}{\ensuremath{\alpha_s(M_Z)^{\overline{MS}}}}
\newc{\alphaimzms}[1]{\ensuremath{\alpha_{#1}(M_Z)^{\overline{MS}}}}
\newc{\alphaemmz}{\ensuremath{\alpha_{\mathrm{em}}(M_Z)^{\overline{MS}}}}

\newc{\mzero}{\ensuremath{{m_0}}}
\newc{\mhalf}{\ensuremath{ m_{1/2}}}
\newc{\tanb}{\ensuremath{\tan\beta}}
\newc{\azero}{\ensuremath{ A_0}}
\newc{\bzero}{\ensuremath{ B_0}}
\newc{\signmu}{\ensuremath{\rm{sgn}\,\mu}}
\newc{\mueff}{\ensuremath{\mu_{\rm{eff}}}}
\newc{\lam}{\ensuremath{{\lambda}}}
\newc{\kap}{\ensuremath{{\kappa}}}
\newc{\alam}{\ensuremath{{A_{\lambda}}}}
\newc{\akap}{\ensuremath{{A_{\kappa}}}}
\newc{\hs}{\ensuremath{ H_s}}      
\newc{\mhs}{\ensuremath{ m_{H_s}}} 
\newc{\mgut}{\ensuremath{ M_{\rm GUT}}}
\newc{\mplanck}{\ensuremath{ M_{\rm P}}}      \newc{\mpl}{\ensuremath{ M_{\rm Pl}}}
\newc{\msusy}{\ensuremath{ M_{\rm SUSY}}}      \newc{\ms}{\ensuremath{ M_{\rm S}}}
 \newc{\mhl}{\ensuremath{m_\hl}} 
 \newc{\mhone}{\ensuremath{m_{h_1}}} 
 \newc{\mhtwo}{\ensuremath{m_{h_2}}} 
 \newc{\mglu}{\ensuremath{m_{\tilde g}}} 
 \newc{\mul}{\ensuremath{m_{\tilde{u}_L}}} 
 \newc{\mtone}{\ensuremath{m_{\tilde{t}_1}}} 
 \newc{\ma}{\ensuremath{m_A}} 
 \newc{\maone}{\ensuremath{m_{a_1}}} 
 \newc{\matwo}{\ensuremath{m_{a_2}}}
 \newc{\hone}{\ensuremath{h_1}}
 \newc{\htwo}{\ensuremath{h_2}}
 \newc{\aone}{\ensuremath{a_1}}
 \newc{\atwo}{\ensuremath{a_2}}
 \newc{\mhu}{\ensuremath{ m_{H_u}}}       
 \newc{\mhd}{\ensuremath{ m_{H_d}}}
 \newc{\mhusq}{\ensuremath{ m_{H_u}^2}}       
 \newc{\mhdsq}{\ensuremath{ m_{H_d}^2}}
 \newc{\mhuew}{\ensuremath{ m^{\ast}_{H_u}}}       
 \newc{\mhdew}{\ensuremath{ m^{\ast}_{H_d}}}
 \newc{\mhuewsq}{\ensuremath{ m^{\ast\, 2}_{H_u}}}       
 \newc{\mhdewsq}{\ensuremath{ m^{\ast\, 2}_{H_d}}}
 \newc{\hu}{\ensuremath{ H_u}}       
 \newc{\hd}{\ensuremath{ H_d}}
 \newc{\barmhu}{\ensuremath{ \bar{m}_{H_u}}}
 \newc{\barmhd}{\ensuremath{ \bar{m}_{H_d}}}

 \newc{\mqthree}{\ensuremath{m_{\widetilde{Q}_3}^2}}
 \newc{\muthree}{\ensuremath{m_{\tilde{u}_3}^2}}
 \newc{\mdthree}{\ensuremath{m_{\tilde{d}_3}^2}}
 \newc{\mlthree}{\ensuremath{m_{\widetilde{L}_3}^2}}
 \newc{\methree}{\ensuremath{m_{\tilde{e}_3}^2}}
 \newc{\mqtwo}{\ensuremath{m_{\widetilde{Q}_2}^2}}
 \newc{\mutwo}{\ensuremath{m_{\tilde{u}_2}^2}}
 \newc{\mdtwo}{\ensuremath{m_{\tilde{d}_2}^2}}
 \newc{\mltwo}{\ensuremath{m_{\widetilde{L}_2}^2}}
 \newc{\metwo}{\ensuremath{m_{\tilde{e}_2}^2}}
 \newc{\mqone}{\ensuremath{m_{\widetilde{Q}_1}^2}}
 \newc{\muone}{\ensuremath{m_{\tilde{u}_1}^2}}
 \newc{\mdone}{\ensuremath{m_{\tilde{d}_1}^2}}
 \newc{\mlone}{\ensuremath{m_{\widetilde{L}_1}^2}}
 \newc{\meone}{\ensuremath{m_{\tilde{e}_1}^2}}
 \newc{\msmul}{\ensuremath{m_{\tilde{\mu}_L}}}
 \newc{\msmur}{\ensuremath{m_{\tilde{\mu}_R}}}
 \newc{\msneumu}{\ensuremath{m_{\tilde{\nu}_{\mu}}}}
 \newc{\mone}{\ensuremath{M_1}}
 \newc{\monesq}{\ensuremath{M_1^2}}
 \newc{\mtwo}{\ensuremath{M_2}}
 \newc{\mtwosq}{\ensuremath{M_2^2}}
 \newc{\mthree}{\ensuremath{M_3}}
 \newc{\mthreesq}{\ensuremath{M_3^2}}
 \newc{\atau}{\ensuremath{{A_{\tau}}}}
 \newc{\at}{\ensuremath{{A_{t}}}}
 \newc{\ab}{\ensuremath{{A_{b}}}}
 \newc{\atausq}{\ensuremath{{A_{\tau}^2}}}
 \newc{\atsq}{\ensuremath{{A_{t}^2}}}
 \newc{\absq}{\ensuremath{{A_{b}^2}}}

 \newc{\dmzero}{\ensuremath{\Delta{_{m_0}}}}
 \newc{\dmhalf}{\ensuremath{\Delta{_{m_{1/2}}}}}
 \newc{\dmu}{\ensuremath{\Delta{_{\mu}}}}

 \newc{\pten}{\ensuremath{\psi_{10}}}
 \newc{\ffive}{\ensuremath{\phi_{5}}}
 \newc{\hfive}{\ensuremath{h_{5}}}
 \newc{\hbfive}{\ensuremath{h_{\bar{5}}}}
 \newc{\thet}{\ensuremath{\theta_{50}}}
 \newc{\thetb}{\ensuremath{\theta_{\,\overline{50}}}}
 \newc{\ptenhat}{\ensuremath{\hat{\psi}_{10}}}
 \newc{\ffivehat}{\ensuremath{\hat{\phi}_{5}}}
 \newc{\hfivehat}{\ensuremath{\hat{h}_{5}}}
 \newc{\hbfivehat}{\ensuremath{\hat{h}_{\bar{5}}}}
 \newc{\thethat}{\ensuremath{\hat{\theta}_{50}}}
 \newc{\thetbhat}{\ensuremath{\hat{\theta}_{\,\overline{50}}}}
 \newc{\si}{\ensuremath{\Sigma}}
 \newc{\mfive}{\ensuremath{m_5^2}}
 \newc{\mten}{\ensuremath{m_{10}^2}}
 \newc{\dfive}{\ensuremath{\Delta^2_5}}
 \newc{\dbfive}{\ensuremath{\Delta^2_{\bar{5}}}}
 \newc{\dfifty}{\ensuremath{\Delta^2_{50}}}
 \newc{\dfiftyb}{\ensuremath{\Delta^2_{\,\overline{50}}}}
 \newc{\msi}{\ensuremath{m_{\Sigma}^2}}
 \newc{\lamh}{\ensuremath{\lambda_{H}}}
 \newc{\lamhb}{\ensuremath{\lambda_{\bar{H}}}}
 \newc{\ah}{\ensuremath{A_{H}}}
 \newc{\ahb}{\ensuremath{A_{\bar{H}}}}
 \newc{\lams}{\ensuremath{\lambda_{S}}}
 \newc{\as}{\ensuremath{A_{S}}}
 \newc{\lamsig}{\ensuremath{\lambda_{\si}}}
 \newc{\asig}{\ensuremath{A_{\si}}}

 \newc{\msten}{\ensuremath{m_{16}^2}}
 \newc{\mhun}{\ensuremath{m_{126}^2}}
 \newc{\mhunb}{\ensuremath{m_{\bar{126}}^2}}
 \newc{\mthun}{\ensuremath{m_{210}^2}}
 \newc{\ahun}{\ensuremath{A_{\bar{126}}}}
 \newc{\yhun}{\ensuremath{Y_{\bar{126}}}}
 \newc{\aten}{\ensuremath{A_{10}}}
 \newc{\yten}{\ensuremath{Y_{10}}}
 \newc{\alone}{\ensuremath{A_{\lambda_1}}}
 \newc{\altwo}{\ensuremath{A_{\lambda_2}}}
 \newc{\althree}{\ensuremath{A_{\lambda_3}}}
 \newc{\althreeb}{\ensuremath{A_{\bar{\lambda_3}}}}
 \newc{\lone}{\ensuremath{\lambda_1}}
 \newc{\ltwo}{\ensuremath{\lambda_2}}
 \newc{\lthree}{\ensuremath{\lambda_3}}
 \newc{\lthreeb}{\ensuremath{\bar{\lambda_3}}}

\newc{\sigsip}{\ensuremath{\sigma^{\rm SI}_{p}}}	\newc{\sigsin}{\ensuremath{\sigma^{\rm SI}_{n}}}
\newc{\sigsdp}{\ensuremath{\sigma^{\rm SD}_{p}}}	\newc{\sigsdn}{\ensuremath{\sigma^{\rm SD}_{n}}}
\newc{\sigsi}{\ensuremath{\sigma^{\rm SI}}}	\newc{\sigsd}{\ensuremath{\sigma^{\rm SD}}}
\newc{\sigv}{\ensuremath{\sigma v}}
\newc{\abund}{\ensuremath{ \Omega h^2}}
\newc{\omegadm}{\ensuremath{ \Omega_{{\rm DM}}}}     \newc{\abunddm}{\ensuremath{ \Omega_{{\rm DM}} h^2}} 
\newc{\omegam}{\ensuremath{ \Omega_{{\rm m}}}}       \newc{\abundm}{\ensuremath{ \Omega_{{\rm m}} h^2}}
\newc{\omegab}{\ensuremath{ \Omega_{{\rm b}}}}	\newc{\abundb}{\ensuremath{ \Omega_{{\rm b}} h^2}}
\newc{\omegatot}{\ensuremath{ \Omega_{{\rm TOT}}}}
\newc{\omegacdm}{\ensuremath{ \Omega_{{\rm CDM}}}}   \newc{\abundcdm}{\ensuremath{ \Omega_{{\rm CDM}} h^2}}
\newc{\omegalambda}{\ensuremath{ \Omega_{\Lambda}}} \newc{\abundlambda}{\ensuremath{ \Omega_{\Lambda} h^2}}
\newc{\omegarad}{\ensuremath{ \Omega_{{\rm rad}}}}  \newc{\abundrad}{\ensuremath{ \Omega_{{\rm rad}} h^2}}
\newc{\rhocrit}{\ensuremath{ \rho_{\rm crit}}}
\newc{\rhochi}{\ensuremath{ \rho_{\chi}}}
\newc{\abunchi}{\ensuremath{\Omega_\chi h^2}}
\newc{\abundlsp}{\ensuremath{\Omega_{\rm LSP}h^2}}

\newc{\amu}{\ensuremath{ a_{\mu}}}        \newc{\amususy}{\ensuremath{ a_{\mu}^{\mathrm{SUSY}}}}
\newc{\amuexpt}{\ensuremath{ a_{\mu}^{\mathrm{expt}}}}        \newc{\amusm}{\ensuremath{ a_{\mu}^{\mathrm{SM}}}}
\newc\deltaamu{\ensuremath{\Delta a_{\mu}}} \newc{\deltaamususy}{\ensuremath{\delta a_{\mu}^{\mathrm{SUSY}}}}
\newc\gmtwo{\ensuremath{ (g-2)_{\mu}}} 
\newc{\deltagmtwomususy}{\ensuremath{\delta\left(g-2\right)_{\mu}^{\mathrm{SUSY}}}}
\newc{\deltagmtwomu}{\ensuremath{\delta\left(g-2\right)_{\mu}}}
\newc\BR{\ensuremath{\rm BR}}
\newc\bsgamma{\ensuremath{ b\rightarrow s \gamma }}
\newc\bxsgamma{\ensuremath{\overline{B}\rightarrow X_{s}\gamma}}
\newc\brbsgamma{\ensuremath{\BR\left(\bsgamma\right)}}
\newc\brbxsgamma{\ensuremath{\BR\left(\bxsgamma\right)}}
\newc\bsmumu{\ensuremath{B_s\to\mu^+\mu^-}}
\newc\brbsmumu{\ensuremath{\BR\left(B_s\to\mu^+\mu^-\right)}}
\newc\bdmmumu{\ensuremath{\overline{B}_d\to\mu^+\mu^-}}
\newc\bbbarmix{\ensuremath{\overline{B}_s\mbox{-}B_s}}      
\newc\delmbs{\ensuremath{\Delta M_{B_s}}}
\newc{\butaunu}{\ensuremath{B_u \rightarrow \tau \nu}}
\newc{\brbutaunu}{\ensuremath{\BR\left(B_u \rightarrow \tau \nu\right)}}

\newcommand*{\reffig}[1]{Fig.~\ref{#1}}
 
        \newcommand*{\refeq}[1]{Eq.~(\ref{#1})}
     \newcommand*{\refsec}[1]{Sec.~\ref{#1}}

\newcommand*{\pythia}{\text{PYTHIA}}

\let\oldcite\cite
\renewcommand*{\cite}{~\oldcite}

\newcommand*{\hl}{\ensuremath{h}}

\newcommand*{\madgr}{\texttt{MadGraph5\_aMC@NLO}}

\restylefloat{figure}

\title{Less-simplified models of dark matter for direct detection and the LHC}

\author[a]{Arghya Choudhury,}
\author[b]{Kamila Kowalska,}
\author[b]{Leszek Roszkowski,\footnote{On leave of absence from the University of Sheffield, U.K.}}
\author[b]{Enrico Maria Sessolo}
\author[b]{and Andrew J. Williams}

\affiliation{$^a$ Regional Centre for Accelerator-based Particle Physics,
Harish-Chandra Research Institute,\\ Allahabad - 211019, India\\
$^b$ National Centre for Nuclear Research,\\
  Ho{\. z}a 69, 00-681 Warsaw, Poland} 

\emailAdd{arghyachoudhury@hri.res.in}
\emailAdd{kamila.kowalska@ncbj.gov.pl}
\emailAdd{L.Roszkowski@sheffield.ac.uk}
\emailAdd{enrico.sessolo@ncbj.gov.pl}
\emailAdd{andrew.williams@ncbj.gov.pl}

\abstract{We construct models of dark matter with suppressed spin-independent scattering cross section utilizing the existing simplified model framework.
Even simple combinations of simplified models can exhibit interference effects that cause the tree level contribution to the scattering cross section to vanish, 
thus demonstrating that direct detection limits on simplified models are not robust when embedded in a more complicated and realistic framework. 
In general for fermionic WIMP masses $\gsim 10\gev$ 
direct detection limits on the spin-independent scattering cross section are much stronger than those coming from the LHC.
However these model combinations, which we call less-simplified models, 
represent situations where LHC searches become more competitive than direct detection experiments even for moderate dark matter mass.
We show that a complementary use of several searches at the LHC can strongly constrain the direct detection blind spots by setting limits on 
the coupling constants and mediators' mass.
We derive the strongest limits for combinations of vector + scalar, vector + ``squark", and ``squark" + scalar mediator,
and present the corresponding projections for the LHC 14\tev\ for 
a number of searches: mono-jet, jets + missing energy, and searches for heavy vector resonances.}

\begin{document}
\maketitle
\section{\label{sec:intro}Introduction}

In the past few years simplified model spectra 
(SMS)\cite{Goodman:2011jq,Abdallah:2014hon,Malik:2014ggr,Abdallah:2015ter,Abercrombie:2015wmb} have been used 
as a reasonable framework for the interpretation 
of the limits from mono-jet/mono-photon\cite{Aad:2013oja,Aad:2014vka,Aad:2014tda,Aad:2015zva,
Khachatryan:2014rra,Khachatryan:2014rwa} searches on direct production of dark matter (DM) at the LHC.
In SMS one generally parametrizes the cross section in terms of a few parameters, like the couplings of the DM 
with the visible sector, or the mass of the particles assumed to mediate the 
interaction between the DM and the partons in the nucleons, effectively obtaining 
a way to capture the physical properties of a vast class of weakly interacting massive particle (WIMP) scenarios without excessive 
proliferation in the number of free parameters.

Early LHC results presented by the experimental collaborations were often interpreted in the effective field theory (EFT) 
framework\cite{Cao:2009uw,Beltran:2010ww,Goodman:2010yf,Bai:2010hh,Goodman:2010qn,Rajaraman:2011wf,Fox:2011pm,Cheung:2012gi,Matsumoto:2014rxa}, which 
gives a good approximation as long as the interaction is mediated by particles with masses well above the collision energy. 
The EFT framework also has the advantage of providing bounds in terms of a common contact operator that can be used for 
comparison with the limits on the spin-independent DM-nucleon cross section, \sigsip, 
obtained in underground experiments like XENON100\cite{Aprile:2012nq} and LUX\cite{Akerib:2013tjd}, 
or the spin-dependent cross section, \sigsdp, measured for example at PICASSO\cite{BarnabeHeider:2005pg,Archambault:2012pm}, COUPP\cite{Behnke:2012ys} 
and, indirectly, IceCube\cite{IceCube:2011aj,Aartsen:2012kia} and ANTARES\cite{Adrian-Martinez:2013ayv}.

However, at the center-of-mass energies typically probed in a collider environment
it is often necessary to consider situations where the approximations underlying the 
EFT framework no longer apply\cite{Fox:2011fx,Shoemaker:2011vi,Busoni:2013lha,Busoni:2014sya,Busoni:2014haa,Racco:2015dxa}.
Several recent analyses that compared LHC and direct detection (DD) bounds on DM have therefore rather used the SMS 
framework\cite{An:2012va,Frandsen:2012rk,An:2013xka,Bai:2013iqa,DiFranzo:2013vra,Alves:2013tqa,Papucci:2014iwa,Garny:2014waa,
Buchmueller:2014yoa,Busoni:2014gta,
Buckley:2014fba,Garny:2015wea,Chala:2015ama,Duerr:2015wfa}.
A wide range of possibilities for the DM mediators and couplings has been discussed 
in\cite{An:2012va,Frandsen:2012rk,An:2013xka,Bai:2013iqa,DiFranzo:2013vra,Alves:2013tqa,Papucci:2014iwa,Garny:2014waa,
Buchmueller:2014yoa,Busoni:2014gta,
Buckley:2014fba,Garny:2015wea,Chala:2015ama,Duerr:2015wfa}, 
but in general the literature agrees in that for WIMP masses above $\sim 5\gev$ the bounds on \sigsip\ coming from reinterpretations of mono-jet searches are 
not competitive with the bounds from LUX or XENON100.

On the other hand, because of their reduced number of parameters, 
the most common SMS are not intrinsically equipped to capture some of the interesting phenomenology of more realistic 
theoretical DM models (for example those emerging in low scale supersymmetry (SUSY)).
In models that involve a richer spectrum of particles,  
several effects can arise which are missed in the most simple SMS, like long decay chains, or 
the well known fact that interference between different diagrams, or cancellations in the couplings can give rise to ``blind spots" 
for DD searches\cite{Cohen:2011ec,Cheung:2012qy,Cheung:2013dua,Hill:2013hoa,Huang:2014xua,Crivellin:2015bva,Calibbi:2015nha}.
It is also known that with a larger number of particles with differentiated properties one can make a more effective use of the 
complementarity of different experimental strategies, which can be employed in
combination\cite{Papucci:2014iwa,Arbey:2013iza,Baek:2014jga,Arbey:2015hca}. 

In this regard, then, it would perhaps be interesting to give a detailed look at the detection issues arising in 
cases when one moves just one step beyond the SMS approach, i.e., 
when one tries to build models that are halfway in between those SMS characterized 
by just one type of mediator and interaction mechanism, and a UV complete model.
We take this approach in this paper, in which we combine existing SMS in pairs,
with the goal to somewhat mimic the behavior of a developed UV theory without at the same time 
drastically increasing the number of parameters, or including the full spectrum of a specific model.
We only consider SMS and parameter ranges for which the spin-independent scattering cross section is substantial,
so that a comparison between 
the limits from DD and the LHC is always possible. The combinations involve three popular SMS characterized by vector
mediators, scalar mediators, and colored scalar mediators, and we take the DM particle to be a Dirac fermion. 
The models given here represent 
only a few motivated examples of the many combinations that can be constructed.
We dedicate special attention to the blind spots for DD, which stem 
from interference effects among different diagrams. We show that each 
of the emerging blind spots can be tested in Run 2 of the LHC, by different
experimental strategies. Our study is thus complementary to previous studies in this direction\cite{Cohen:2011ec,Cheung:2012qy,Cheung:2013dua,Hill:2013hoa,Huang:2014xua,Crivellin:2015bva,Calibbi:2015nha}, although, 
to the best of our knowledge, the combinations we consider here have not been analyzed before in this setting.

The paper is organized as follows. In \refsec{sec:models} we review the three known SMS of DM 
that constitute the building blocks from which we construct new models for DM searches.
In \refsec{sec:combos} we present the parameter space of these new models and confront them with the current bounds from searches for DM, which include underground detectors
and DM searches at the LHC. We further identify the regions that give rise to suppressions in \sigsip, and apply to the resulting blind spots
existing bounds from the 8\tev\ and 13\tev\ LHC. We also calculate the corresponding projected reach of the 14\tev\ run, showing that this can significantly 
probe these new regions of the parameter space. 
We finally present our summary and concluding remarks in \refsec{sec:summary}.
 
\section{The model blocks\label{sec:models}}

In this paper we present a phenomenological analysis of a few models of DM 
that should provide ``less simplified" model frameworks (LSMS) that to some extent mimic the characteristics of more generic UV theories.
\medskip

\textbf{Model 1.} Combining vector and Higgs portal mediators;

\textbf{Model 2.} Combining Higgs portal and $t$-channel mediators charged under color;

\textbf{Model 3.} Combining vector and $t$-channel mediators charged under color.
\medskip

\noindent In all three cases we take the DM particle to be a Dirac fermion singlet under the SM symmetries. 

The combinations given above are built out of well known SMS, which we call the model blocks, 
whose characteristics we briefly recall in the next subsections. The model blocks 
should respect the gauge symmetries of the SM and not violate minimal flavor violation (MFV)\cite{Abdallah:2015ter,Baek:2015lna}.
These requirements constrain the allowed forms of the models and their parameters, in contrast to the most general forms\cite{Abdallah:2014hon}. 

\subsection{Vector mediator\label{sec:vecmed}}

The mediator considered is a leptophobic $Z'$. Dark matter SMS based on vector mediators of this kind where studied, e.g.,
in\cite{Dudas:2009uq,An:2012va,Frandsen:2012rk,Dreiner:2013vla,Buchmueller:2013dya,Alves:2013tqa,Buchmueller:2014yoa,Busoni:2014gta,
Jacques:2015zha,Chala:2015ama,Autran:2015mfa,Duerr:2015wfa,Alves:2015dya}.
The Dirac fermion singlet DM particle, $\chi$, is coupled to the new gauge boson, $Z'$. 
The new mediator is assumed to have negligible mixing with the $Z$ boson of the SM,
and to not couple to the SM leptons, so that one can easily evade the strong limits 
from di-lepton resonances at the LHC\cite{Aad:2014cka,Khachatryan:2014fba}.
Note that this assumption will not affect the DD constraints
and the results of the next sections will remain general. 
We also always assume that possible anomalies are canceled by new heavy states above $m_{Z'}$,
which do not contribute to phenomenology at the LHC.\footnote{This assumption is not always warranted, see for instance Ref.\cite{Dobrescu:2015asa}.}

The interaction terms relative to DM detection at the LHC and in underground detectors are 
\be\label{veclagr}
{\cal L} \supset Z'_\mu \bar{\chi}\gamma^\mu (g_\chi^V - g_\chi^A \gamma_5) \chi + \sum_i Z'_\mu \bar{q}_i\gamma^\mu (g_q^V - g_q^A \gamma_5) q_i\,,
\ee
where the index $i$ runs over the quarks and we have universal vector (axial-vector) quark couplings $g_q^V$ ($g_q^A$).
The corresponding vector and axial-vector couplings to the DM are $g_\chi^V$ and $g_\chi^A$, respectively.

In this paper we limit ourselves to the case where WIMPs are produced at the LHC through an on-shell mediator:  $2\mchi < m_{Z'}$\,. 
In this regime the production cross section and mediator width are largely independent of the spin structure
of the couplings\cite{Abercrombie:2015wmb}, so that we can set either $g_{\chi/q}^V$ or $g_{\chi/q}^A$ to zero without loss in generality.
We limit ourselves here to cases of vector boson exchange, which give contribution to the spin-independent WIMP-nucleon scattering cross section,
\sigsip.

We are thus left with 4 parameters for this simplified model,
\be
\left\{\mchi, m_{Z'}, g_{\chi}^V, g_{q}^V\right\},\label{zprime}
\ee
where \mchi\ is the WIMP mass and $m_{Z'}$ the $Z'$ mediator mass. 
This number can be reduced to 3 when $g^{V}_{\chi} = g^{V}_q$, since only the product of the two coupling constants matters for \sigsip.

\subsection{Scalar mediator/Higgs portal}

The second building block is a model such that the fermion DM singlet, $\chi$, is coupled to a new singlet real scalar, $s$. 
This model has been analyzed, e.g., in\cite{Baek:2015lna,Baek:2011aa,Baek:2012uj}, but scalar mediators have also been studied 
in, e.g.,\cite{Ghorbani:2014qpa,Buckley:2014fba,Ghorbani:2014gka,Garny:2015wea,Cotta:2013jna,Fischer:2013hwa,Berlin:2015wwa,Varzielas:2015joa}. 
We note that there are many variations with 
scalar mediators or extended Higgs sectors but this model allows us to keep the DM as a SM singlet and allows combinations with 
the other model blocks.

The terms in the Lagrangian relevant to DM searches are
\be\label{hportal}
{\cal L} \supset - y_\chi \bar{\chi}\chi s - \mu_s s |\Phi|^2 - \lambda_s s^2 |\Phi|^2,
\ee
where $y_\chi$ is the Yukawa coupling between the DM and the singlet, and 
$\mu_s$ is a mass term that induces mixing between $s$ and the SM Higgs doublet, $\Phi$, that gives rise to the 
Higgs boson after electroweak (EW) symmetry breaking. In \refeq{hportal} we neglect eventual 
polynomial terms of the singlet only, which are not important for DM SMS (at least at the tree level), 
and which we assume are fixed by the UV completion.
We assume that $\Phi$ develops the SM vacuum expectation value (vev), $v$: $\Phi\rightarrow 1/\sqrt{2}\,(0,v+h)^{T}$\,, which can be determined in terms
of the SM mass and quartic couplings. 

The $\mu_s$ and $\lambda_s$ Lagrangian terms produce an off-diagonal component in the $(h,s)$ mass matrix. 
The mass matrix is diagonalized by a mixing matrix parametrized by a mixing angle $\theta$,
\be
\left(
\begin{array}{c}
 h_{\textrm{SM}}  \\
 H  \\
\end{array}
\right) = \left(
\begin{array}{cc}
 \cos\theta & \sin\theta  \\
  -\sin\theta & \cos\theta \\
\end{array}
\right)\left(\begin{array}{c}
 h  \\
 s  \\
\end{array}
\right).
\ee
In general we will identify the lightest scalar with the observed SM-like Higgs. 
Higgs physics measurement, EW precision tests, and vacuum stability then constrain $|\sin\theta|$\cite{Farzinnia:2013pga,Belanger:2013kya,Pruna:2013bma,Robens:2015gla}.

After diagonalization the relevant terms in the Lagrangian for DM phenomenology are
\be\label{higport}
{\cal L} \supset -y_\chi \left(h_{\textrm{SM}}\sin\theta+ H\cos\theta\right)\bar{\chi}\chi - 
\frac{1}{\sqrt{2}} \left(h_{\textrm{SM}}\cos\theta - H\sin\theta \right) \sum_f y_f \bar{f}f\,,
\ee
where $y_f$ are the SM Yukawa couplings and $f,\bar{f}$ SM fermions. 
This results in the presence of a heavy scalar mediator, $H$, as well as the SM Higgs, $h_{SM}$, 
that couple the DM to the quarks.
Note that the quarks couple to both mediators in proportion to their Yukawa couplings, as a result of the mixing, automatically in agreement with MFV.
In the spirit of phenomenology one can trade the parameters $\lambda_s$ and $\mu_s$ for 
$\theta$ and the mass of the heavy scalar $m_H$ to produce a simplified model of DM. Note that we implicitly assume that the SM and singlet polynomial terms neglected
in \refeq{hportal} are such that the ratio of the Higgs vevs, $\tan\beta\equiv v/v_s$, is much smaller than 1,
so to be comfortably inside the bounds from perturbative unitarity\cite{Robens:2015gla}.

The DM simplified model is finally described by 4 parameters,
\be
\left\{ m_\chi, m_{H}, \sin 2\theta, y_{\chi}\right\}\,.
\ee

In the next section we will consider, for example, a case where the Higgs portal described here interferes 
with the model of \refsec{sec:vecmed}. However, we foreshadow this discussion by reminding the reader that there is also interference within the 
Higgs portal model itself, as recently pointed out in Ref.\cite{Baek:2015lna}.
Thus, the contributions to \sigsip\ intrinsic to this model are also going to be affected by interference effects.  

\begin{figure}[t]
\centering
\includegraphics[width=0.45\textwidth]{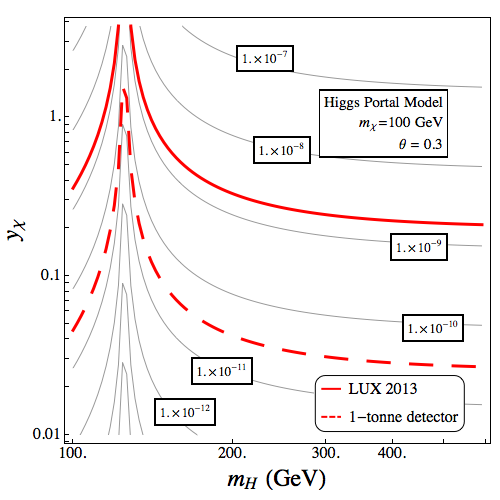}
\caption{Contours of equal spin-independent scattering cross section, \sigsip, 
in pb in the ($m_H$, $y_\chi$) plane for the Higgs portal model. We set $\theta = 0.3$ and $m_{\chi} = 100\gev$. 
The solid red line represents the current 90\%~C.L. upper bound from LUX\cite{Akerib:2013tjd}. 
The dashed red line gives the expected sensitivity of XENON-1T for 2017 or so\cite{Aprile:2012zx}, 
which we use as an example of the generic reach of tonne-scale detectors\cite{Malling:2011va,Amaudruz:2014nsa,Cao:2014jsa}.}
\label{fig:H_h}
\end{figure}

We show this in \reffig{fig:H_h}, where we present contours of \sigsip\ 
in pb in the ($m_H$, $y_\chi$) plane for a fixed $\theta=0.3$\,\footnote{\label{foot1} 
In the analysis of \refsec{sec:combos} we will most often assume 
$\theta=0.2$, which is allowed by the bounds from 
Higgs searches at the LHC in the range $m_H>125\gev$\cite{Robens:2015gla}, by perturbativity of the couplings, and one-loop corrections 
to the $W$ mass\cite{Lopez-Val:2014jva,Robens:2015gla}. As pertains to \reffig{fig:H_h}, one should note that 
$\theta=0.3$ is also excluded for $m_H<125\gev$ by searches at LEP and the LHC\cite{Robens:2015gla}, and that 
the figure is here merely intended as demonstrative of the effects of interference on the WIMP scattering cross section.} and $m_\chi = 100\gev$. 
The cross section depends on $m_\chi$ through the reduced mass of the DM proton system and is thus mostly insensitive to changes in 
$m_\chi$ when $m_\chi \gg m_p$. The current upper bound on \sigsip\ from LUX\cite{Akerib:2013tjd} is shown as a solid red line. 
It is expected to be improved by future tonne-scale underground detectors\cite{Aprile:2012zx,Malling:2011va,Amaudruz:2014nsa,Cao:2014jsa}, 
whose reach is shown here as a dashed red line. 

In general the effective interaction between the quarks and the DM will grow with the mixing angle, the coupling strength $y_\chi$,
and the mass of the heavy scalar, albeit very mildly when $m_{H}\gg m_{h_{\textrm{SM}}}$\,. 
When $m_H \approx m_{h_{\textrm{SM}}}$ the contributions due to $h_{\textrm{SM}}$ and $H$ cancel out and \sigsip\ is suppressed.
This creates a blind spot for DD, characterized by very little sensitivity for underground experiments.

\subsection{Scalar \boldmath$t$-channel mediators}

For the third building block we consider colored scalar mediators coupling the DM directly to the quarks. 
Since these are exchanged in the $t$-channel for DM production at the LHC they are often called $t$-channel mediators.
This case has been analyzed, for example, in\cite{An:2013xka,Bai:2013iqa,DiFranzo:2013vra,Papucci:2014iwa}. 
The new scalars must be charged under color and flavor, so that we can borrow the  
notation used to describe the squarks in the Minimal Supersymmetric Standard Model (MSSM), $\tilde{q}$, even if our model is not necessarily SUSY based.
We assume universality between the first and second generation for the masses of the ``squarks", 
as well as their couplings to the DM, as required by MFV. 
For simplicity, for the first two generations we further assume that all squark masses and couplings are universal, 
thus neglecting the differences
between the left and right components, and up and down squark fields.  
The squarks of the third generation will instead be decoupled and assigned a large mass to avoid the complications associated 
with dedicated stop and sbottom searches.
Besides, these will not contribute to DD observables.

The interaction terms then become,
\be\label{squarks}
{\cal L} \supset \sum_{i=1,2} g_{\tilde{q}}\left( \tilde{u}^{\dag}_{i,R} \bar{\chi}P_R u_i + \tilde{u}^{\dag}_{i,L} \bar{\chi}P_L u_i + \tilde{d}^{\dag}_{i,R} \bar{\chi}P_R d_i + \tilde{d}^{\dag}_{i,L} \bar{\chi}P_L d_i   \right)+\textrm{ h.c.}\,,
\ee
where $g_{\tilde{q}}$ is the coupling strength, $\tilde{u}_{i,L(R)}$ are the left (right) up-type squarks of the $i$th generation,  
$\tilde{d}_{i,L(R)}$ are the left (right) down-type squarks of the $i$th generation, $u_i$ ($d_i$) are the up (down) quarks of the 
$i$th generation and 
$P_R$ and $P_L$ are the right and left chiral projection operators, respectively. 
We assume that the stability of the DM is protected by a discrete symmetry similar to R-parity.
We repeat that all squarks in \refeq{squarks} 
have the same mass in our approximation. 

This simplified model is described by 3 parameters,
\be
\left\{m_\chi, m_{\tilde{q}}, g_{\tilde{q}}\right\}\,,
\ee  
where $m_{\tilde{q}}$ is the universal squark mass and $g_{\tilde{q}}$ is the universal squark-DM coupling. 
Note that in order to preserve the SM gauge symmetries the squarks will develop the same SM gauge interactions as the quarks as 
happens, e.g., in the MSSM. On the other hand, EW gauge and Higgs couplings will not play a role here. This is different from the MSSM, 
where $g_{\tilde{q}}$ is fixed by neutralino mixing and gauge couplings, while in this scenario it is a free parameter.  

\section{Methodology and analysis of the combined models\label{sec:combos}}

We here confront the LSMS introduced at the beginning of \refsec{sec:models}, Model~1, 2, and 3,
with the bounds from DD of DM in underground detectors 
and a number of results from the LHC: mono-jet searches, searches with jets + missing $E_T$ (MET), 
invisible branching fraction of the Higgs boson, and bounds on new heavy $Z'$ resonances from the $t\bar{t}$ and di-jet
invariant mass distributions. 
We will in particular highlight different strategies that can be used to test the parameter space invisible in DD searches. We also 
calculate projections for the LHC 14\tev\ for each model and compare them to the expected sensitivity of tonne-scale underground detectors.   

Models 1, 2, and 3 are implemented in \texttt{FeynRules}\cite{Alloul:2013bka}, 
which is then used to generate \texttt{CalcHEP}\cite{Belyaev:2012qa} and Universal FeynRules Output (UFO) model files.
The spin-independent cross section is calculated with $\tt micrOMEGAs\ v.4.1.8$\cite{Belanger:2013oya}.
For the mono-jet search we generate events at the LHC with \madgr\cite{Alwall:2014hca} and \texttt{\pythia8}\cite{Sjostrand:2007gs} using MLM matching up to two jets. 
For models with squarks we produce both $\chi\bar{\chi}$+jets, $\chi \tilde{q}$+jets associated
production, and $\tilde{q}\tilde{q}^{\ast}$+jets following the example of\cite{Papucci:2014iwa}. 
To derive exclusion bounds for the mono-jet and jets+MET searches we use 
\texttt{CheckMATE}\cite{Drees:2013wra,Barr:2003rg,Cheng:2008hk,Cacciari:2005hq,Cacciari:2008gp,Cacciari:2011ma,
deFavereau:2013fsa,Lester:1999tx,Read:2002hq} and 
two codes developed by some of us, which were previously used in 
Refs.\cite{Kowalska:2015zja,Chakraborti:2014gea,Chakraborti:2015mra}.
The LHC 14\tev\ projections for the mono-jet and jets+MET searches are obtained by implementing the experimental cuts described in\cite{ATL-PHYS-PUB-2014-007} and\cite{ATL-PHYS-PUB-2014-010} 
in our own codes and \texttt{CheckMATE}.
We do not consider other mono-X searches beyond the mono-jet to keep the number of searches considered to a minimum and since projections for 14\tev\ are not currently available.

For the direct searches for $Z'$ we use \madgr\ to compare production cross section 
times branching ratio for $Z'\rightarrow t\bar{t}$ and $Z'\rightarrow q\bar{q}$ 
to the limits given in\cite{Aad:2015fna,Aad:2014aqa,ATLAS:2015nsi}. 
For the 14\tev\ projections we use\cite{ATL-PHYS-PUB-2013-003}.
For the dijet search we use a combination of the 8\tev\ data for $m_{Z'}<1.5\tev$ and 13\tev\ data for $m_{Z'}\gsim 1.5\tev$. 
In the case of the 13\tev\ search the collaboration has recast the search explicitly in terms of the simplified model considered here, 
giving limits in terms of $g_{q}^V$ and $m_{Z'}$\,. 
The collaborations have not provided a 
projection for 300\invfb\ at $14\tev$ in terms of $Z'\rightarrow q\bar{q}$, so that we use the di-top search throughout the paper for consistency.
Finally, we calculate the partial width of the SM Higgs to DM particles, $\Gamma_{h_{\textrm{SM}}\rightarrow \chi\bar{\chi}}$\,, with \texttt{CalcHEP}
and compare to the limit from\cite{ATLAS-CONF-2015-044} following the method described in\cite{Baek:2014jga}.

We do not apply to our models the bounds from the relic density of DM. 
In a sense, this is equivalent to considering the most general phenomenological case, 
neglecting all possible bias from the particular thermal history of the early Universe.
The reader should bear in mind that from this point of view the models 
analyzed in this section are incomplete. In order to satisfy the constraint from the relic density 
one must consider one or more of the following possibilities: large or non-perturbative couplings; additional particles/interactions; or a non-standard thermal history.     

\subsection{Model 1: combining \boldmath$Z'$ and Higgs portal\label{sec:higz}} 

The first LSMS we consider is motivated by the observation that many UV complete models with a $Z'$ 
also contain an extended scalar sector, see for example\cite{Basso:2012ti}.
Inspired by constructions of this kind, we consider a $Z'$ vector boson associated to a new symmetry $U(1)_X$\,, and 
a hypothetical extended scalar sector that will include, among others, 
a $U(1)_X$-neutral SM singlet field $s$ that couples to the SM Higgs and the DM particle like in \refeq{hportal}.
If all other degrees of freedom are decoupled, the low energy Lagrangian is just 
the sum of \refeq{veclagr} and \refeq{higport}.\footnote{A different combination of the same two mediators is given, e.g., in\cite{Ghorbani:2015baa}.}

As explained above, we consider only the vector couplings to the $Z'$\,. 
If we make the assumption that $g_{\chi}^V=g_{q}^V\equiv g_{\chi/q}^V$
we are left with 6 free parameters,
\be
\left\{ \mchi, m_{Z'}, m_H, \theta, y_\chi, g_{\chi/q}^V\right\}\,.
\ee
We assume mixing as maximal as is allowed by the LHC constraints, perturbativity of the couplings, and EW precision observables: $\theta=0.2$ (see Footnote~\ref{foot1}).  
The cross section \sigsip\ depends mildly on the angle, via $\sin 2\theta$.

For a nuclear element $N$, the differential WIMP-nucleus scattering cross section in the non-relativistic limit is given by\cite{Jungman:1995df}
\be 
\frac{d\sigma_{\chi N}}{d|\mathbf{q}|^2}=\frac{1}{\pi v_{\chi}^2}\left[Z f_p+(A-Z) f_n\right]^2 F^2(Q)\,,\label{cross}
\ee
where $|\mathbf{q}|$ is the transferred momentum, $Z$ is the atomic number, $A$ the atomic weight, $v_{\chi}$ is the average speed of the DM in the halo, and $F(Q)$ is the Wood-Saxon function given in\cite{Jungman:1995df} as a function of $Q=|\mathbf{q}|^2/2 m_{N}$\,.

The contributions to $f_p$ and $f_n$ come from the effective interactions of the WIMP with protons and neutrons, respectively, and they
are approximately given in Model~1 by 
\be
f_n\approx f_p\approx \frac{y_{\chi}\sin 2\theta}{4\,m_{h_{\textrm{SM}}}^2}\left(1-\frac{m_{h_{\textrm{SM}}^2}}{m_H^2}\right)\frac{m_p}{v}\left(\sum_{q=u,d,s}f_{Tq}+\frac{2}{9}\,f_{TG}\right)+\frac{3}{2}\frac{g_{\chi}^Vg_q^V}{m_{Z'}^2}\,,\label{ampl}
\ee 
where $m_p$ is the nucleon mass, and $f_{Tq}$ and $f_{TG}$ are the hadronic matrix elements defined for example in\cite{Jungman:1995df}. 
We adopt in our study their default values embedded in \texttt{micrOMEGAs}: $f_{Td}=0.0191$, $f_{Tu}=0.0153$, $f_{Ts}=0.0447$, and $f_{TG}=1-f_{Tu}-f_{Td}-f_{Ts}$. 

\begin{figure}[t]
\centering
\subfloat[]{%
\label{fig:a}%
\includegraphics[width=0.43\textwidth]{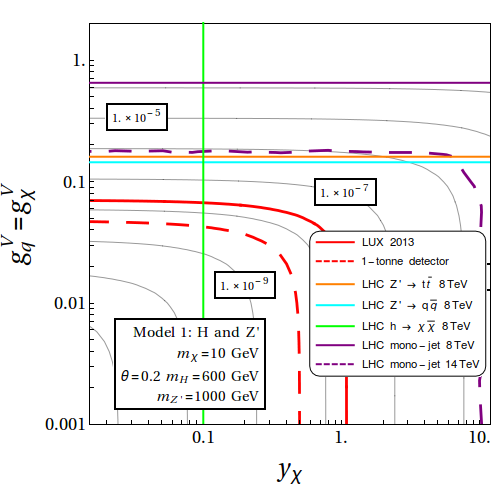}
}%
\hspace{0.07\textwidth}
\subfloat[]{%
\label{fig:b}%
\includegraphics[width=0.43\textwidth]{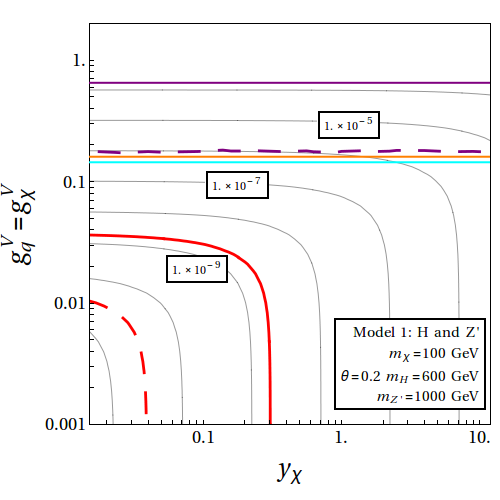}
}%
\hspace{0.07\textwidth}
\subfloat[]{%
\label{fig:c}%
\includegraphics[width=0.43\textwidth]{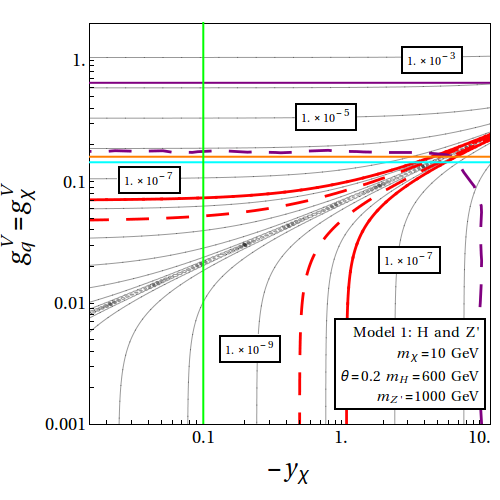}
}%
\hspace{0.07\textwidth}
\subfloat[]{%
\label{fig:d}%
\includegraphics[width=0.43\textwidth]{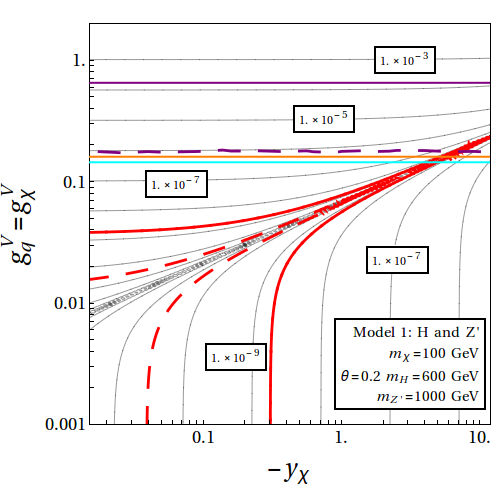}
}%
\caption{(a) Spin-independent scattering cross section in pb in the ($y_\chi$, $g_{\chi/q}^V$) plane for a model combining 
$Z'$ mediator and Higgs portal. We set $\mchi = 10\gev$, $m_{Z'}=1000\gev$, $\theta = 0.2$ and $m_H = 600\gev$. 
The solid red line represents the current 90\%~C.L. upper bound from LUX and the dashed red line is the projected limit for tonne-scale detectors
in 2017 or so. Solid purple line is the 95\%~C.L. upper bound from the 8\tev\ mono-jet search at ATLAS\cite{Aad:2015zva}. 
Dashed purple line is the projected limit from the mono-jet search at 14\tev\ with 300\invfb\cite{ATL-PHYS-PUB-2014-007}. 
Solid orange line gives the upper limit from searches for heavy vector resonances in the di-top channel at 8 TeV\cite{Aad:2015fna}, and the solid cyan line gives the equivalent limit in the $\bar{q}q$ search\cite{Aad:2014aqa}. 
Green solid line gives the upper limit from the invisible width of the Higgs boson in a CMS/ATLAS combined analysis\cite{ATLAS-CONF-2015-044}. 
(b) Same as (a) but $\mchi=100\gev$. (c) Same as (a) but the sign of $y_{\chi}$ is negative. (d) Same as (c) but $\mchi=100\gev$.}
\label{fig:Z_H}
\end{figure}

If $y_{\chi}>0$\,, $m_{H}\gg m_{h_{\textrm{SM}}}$, and $g_{\chi}^V=g_{q}^V$, destructive interference between the terms in \refeq{ampl}
does not take place.   
In \reffig{fig:Z_H}(a) we show contours of \sigsip\ in pb in the 
($y_{\chi}$, $g_{\chi/q}^V$) plane in this case, for fixed values of the DM mass, 
$\mchi=10\gev$, and of the mediators' mass, $m_H = 600\gev$ and $m_{Z'}=1000\gev$.
The mediators' mass has been chosen so to be within the present limits from the LHC for 
reasonable choices of the couplings, not too far from their 
SM strengths, and also so that we are reasonably far away from interference effects between the light and heavy Higgs bosons.

%

As was the case in \reffig{fig:H_h}, the solid red line shows the 90\%~C.L. upper bound on \sigsip\ from LUX  
and the corresponding red dashed line the projected reach of XENON-1T. 
The upper bound from mono-jet searches at the LHC 8\tev\cite{Aad:2015zva} is shown as a solid purple line, 
and the projected reach for mono-jet with 300\invfb\ at 14\tev\cite{ATL-PHYS-PUB-2014-007}
is shown with a dashed purple line.
The solid orange line shows the upper bound on heavy resonances from the invariant $t\bar{t}$ mass distribution after the LHC 8\tev\ 
run\cite{Aad:2015fna}  and the solid cyan line gives the equivalent limit in the $\bar{q}q$ channel\cite{Aad:2014aqa}. 
The projected reach for the $\bar{t}t$ search at the LHC 14\tev\cite{ATL-PHYS-PUB-2013-003} with 300\invfb\ does not improve on the 8\tev\ 
bound for $m_{Z'}=1000\gev$ and is consequently not shown here.
Finally, the green solid vertical line gives the upper limit on the $y_{\chi}$ coupling from a CMS/ATLAS combined analysis of the invisible decay of the 
Higgs boson\cite{ATLAS-CONF-2015-044}.

As is well known, the DD detection bound on $g_{\chi/q}^V$ from LUX is significantly more constraining 
then any of the collider limits, even for a relatively light DM mass. 
LUX loses sensitivity with respect to the 
collider bounds when $\mchi\lesssim5\gev$. 
For $\mchi\lesssim 62\gev$, the invisible width of the Higgs boson analysis places an upper bound on $y_{\chi}$ 
that is stronger than the projected reach of many tonne-scale detectors, as was also recently pointed out in\cite{Baek:2014jga}. 

When the WIMP mass becomes larger than 10\gev\ the DD bounds become more severe, and they reach their maximal sensitivity when 
$\mchi\approx 50\gev$. We show the bounds for $\mchi=100\gev$ in \reffig{fig:Z_H}(b). 
Note that the invisible width bound does not apply to this case, so that the strongest limits on both couplings here come 
from DD experiments. Also, the projected upper bound on $y_{\chi}$ 
from mono-jet searches at the LHC 14\tev\ is stronger when $\mchi\lesssim 62\gev$, because the $\chi\bar{\chi}$ pair can be produced via an on-shell SM Higgs.

On the other hand, if $y_{\chi}<0$\,, or if it is positive but $g_{\chi}^V=-g_{q}^V$, 
the diagrams corresponding to the $Z'$ and Higgs portal interfere destructively and \sigsip\ becomes suppressed, as can be inferred from \refeq{ampl}.
Note, incidentally, that the cancellation can only happen in the nonrelativistic limit. To see this one can consider the relativistic 
WIMP-quark scattering, $q(p_1)\chi(p_3)\rightarrow q(p_2)\chi(p_4)$, for which the squared amplitude reads: 
\bea
|\mathcal{A}|^2&=&2\,\frac{\sin^2 2\theta\,y_q^2 y_{\chi}^2 (m_p^2+p_1p_2)(m_\chi^2+p_3p_4)}{[(p_1-p_2)^2-m_{h_{\textrm{SM}}}^2]^2}+\frac{(g_{\chi}^Vg_q^V)^2\,(16m_p^2-8p_1p_2)(16m_{\chi}^2-8p_3p_4)}{[(p_1-p_2)^2-m_{Z'}^2]^2}
\nonumber\\
 &+&\frac{16}{\sqrt{2}}\frac{\sin 2\theta\,y_q y_{\chi} g_{\chi}^V g_q^V m_p m_\chi (p_1+p_2)^\mu(p_3+p_4)_\mu}{[(p_1-p_2)^2-m_{h_{\textrm{SM}}}^2][(p_1-p_2)^2-m_{Z'}^2]}\,.\label{relat}
\eea
Obviously, without a detailed knowledge of the incoming and outgoing 4-momenta, a cancellation between the terms of 
\refeq{relat} cannot take place. Additionally, in \refeq{ampl} there is also an effective term due to the coupling of the Higgs to the gluon field strength, which determines the position of 
the blind spot.

The effects of the cancellation are shown for $\mchi=10\gev$ in \reffig{fig:Z_H}(c). 
The blind spot is in the plots a narrow diagonal region, over which the value of \sigsip\ visibly drops below
the potential reach of tonne-scale detectors.  
The corresponding case for $\mchi=100\gev$ is shown in \reffig{fig:Z_H}(d).

From \refeq{ampl} one can derive the condition for the blind spot:
\be 
y_{\chi}\approx -\left(\frac{8.22\times 10^7\textrm{ GeV}^2}{m_{Z'}^2}\right)
\frac{g_{\chi}^V g_{q}^V}{\sin 2\theta\, \left(1-\frac{m_{h_{\textrm{SM}}}^2}{m_H^2}\right)}\,.\label{BS1}
\ee
Condition~(\ref{BS1}) shows that the contributions to the amplitude of the diagrams from the $Z'$ and Higgs portal are of comparable size 
for comparable coupling strengths if $m_{Z'}$ is at least of the order of a \tev\ or larger.
Here we neglect further loop corrections which may reintroduce a suppressed coupling to the nucleus.
When \refeq{BS1} is satisfied the model is beyond the reach of DD searches in underground detectors 
but can be tested by collider means.

\begin{figure}[t]
\centering
\subfloat[]{%
\label{fig:a}%
\includegraphics[width=0.45\textwidth]{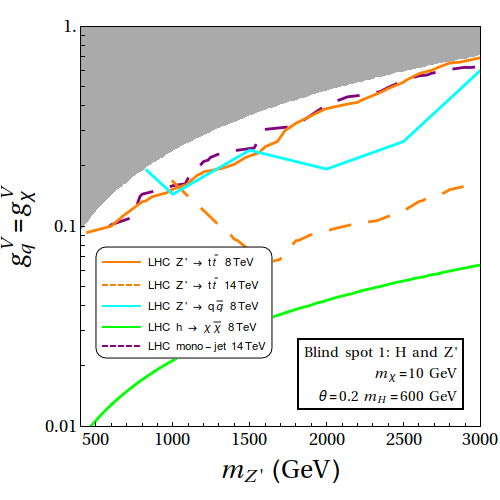}
}
\hspace{0.07\textwidth}
\subfloat[]{%
\label{fig:c}%
\includegraphics[width=0.45\textwidth]{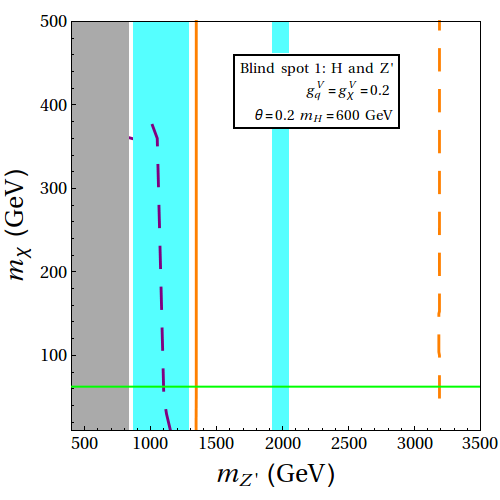}
}%
\caption{(a) Interplay of LHC constraints for the blind spot parametrized by \refeq{BS1} in the ($m_{Z'}$, $g_{\chi/q}^V$) plane. Here $\mchi=10\gev$, $m_{H}=600\gev$, and $\theta=0.2$. 
Solid orange line gives the 95\%~C.L. upper bound from the ATLAS search for heavy resonances in the di-top channel at 8\tev. 
The dashed orange line gives the corresponding reach of the LHC 14\tev\ with 300\invfb\cite{ATL-PHYS-PUB-2013-003}. 
Solid cyan line gives a combination of the upper bounds on the $Z'$ invariant mass in $\bar{q}q$ searches at 8\tev\cite{Aad:2014aqa} and 13\tev\cite{ATLAS:2015nsi}.
Green solid line gives the combined ATLAS/CMS upper bound from the invisible width of the Higgs. Dashed purple line gives the projected mono-jet reach at ATLAS with 14\tev\ and 300\invfb. 
(b) The bounds projected to the ($m_{Z'}$, \mchi) plane for $g_{\chi/q}^V=0.2$ and $m_{H}=600\gev$. 
Corresponding bounds are now lower limits on the masses. The combination of 8 and 13\tev\ data in $\bar{q}q$ searches excludes the parameter space within cyan bars.}
\label{fig:Z_H_det}
\end{figure}

We therefore show in what follows how the interplay of mono-jet searches, searches for $Z'$ resonances, and Higgs width measurements 
introduced above constrains the remaining parameters of Model~1, when the condition (\ref{BS1}) for a blind spot is satisfied.  
In \reffig{fig:Z_H_det}(a) the bounds are shown in the ($m_{Z'}$, $g_{\chi/q}^V$) plane, assuming \refeq{BS1} holds.
The DM mass is here fixed, $\mchi=10\gev$, and we have set $m_{H}=600\gev$. 
Note that the plots essentially do not change much for any $m_{H} \gsim 200\gev$.

The shaded region at the top of \reffig{fig:Z_H_det}(a) is not allowed, as $y_{\chi}$ becomes there nonperturbative, $y_{\chi}> 4\pi$. 
In both panels the color code is the same as in \reffig{fig:Z_H}. Note that the strongest upper bound is currently 
given by the invisible width of the Higgs,
but one must remember that the bound does not apply for $\mchi\gsim 62\gev$. 
The limits from the mono-jet and direct $Z'$ resonance searches, on the other hand, barely change position
over a large range of DM masses. 
The current ATLAS mono-jet bound is too weak 
and does not appear in \reffig{fig:Z_H_det}(a), in agreement with the results of Figs.~\ref{fig:Z_H}(c) and \ref{fig:Z_H}(d). 
The upper limit from $Z'\rightarrow \bar{t}t$ is shown with a solid orange line and a combination of the upper bounds on the $Z'$ invariant mass 
in $\bar{q}q$ searches at 8\tev\cite{Aad:2014aqa} and 13\tev\cite{ATLAS:2015nsi} is shown with a cyan solid line. 
Note that the 
13\tev\ data improved significantly on the 8\tev\ data 
for $Z'$ masses above 1500\gev. 

We calculate that in Run 2 with 300\invfb\ the bounds from resonance searches will improve considerably for large 
masses, as shown by the orange dashed line.
When $m_{Z'}\lesssim 1500\gev$ the improvement over the 8\tev\ bound is small and for $m_{Z'}\approx 1000\gev$ we project that there will be no improvement in 
the $\bar{t}t$ resonance search, which explains why the orange dashed line does not appear in \reffig{fig:Z_H}.
Note that the 14\tev\ mono-jet reach is comparable to the current di-top bound at 8\tev\ and far below the 14\tev\ 
projection making mono-jet searches less efficient than other strategies for testing this model.

The mono-jet bounds derived above feature just a mild dependence on the DM mass, since the probability for emission of a hard ISR jet is largely independent of it, as long as
the mass is small compared to the cut on the transverse momentum of the leading jet.
In \reffig{fig:Z_H_det}(b) we show the parameter space that can by probed by mono-jet searches in the ($m_{Z'}$, \mchi) plane for $g_{\chi/q}^V=0.2$.
It can be seen that the mono-jet search is always weaker than the direct search for a $Z'$ in the di-top and di-jet channels.
The 14\tev\ mono-jet search loses all sensitivity once the DM mass becomes of the order of $\sim350\gev$, as expected for the couplings and selection
requirements considered here.
     
\begin{figure}[t]
\centering
\includegraphics[width=0.45\textwidth]{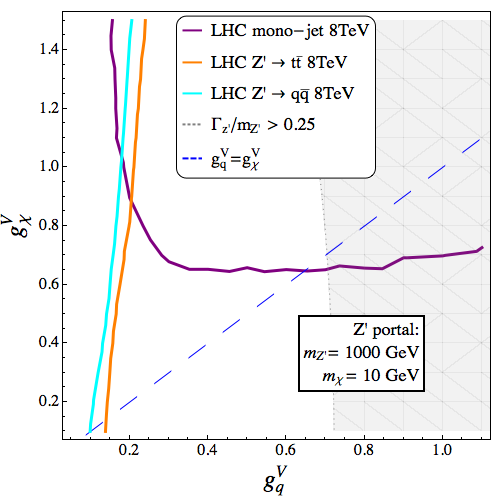}
\caption{A comparison of the limits from mono-jet searches at the LHC and searches for $Z'$ in the di-top and di-jet invariant mass distributions
in the ($g_{q}^V$, $g_{\chi}^V$) plane for $m_{Z'}=1000\gev$, $\mchi=10\gev$.}
\label{fig:gqgx}
\end{figure}

Finally, let us conclude this subsection with a few more comments. 
Given particular values for the $Z'$ mediator mass, 
the limits shown in the plots are derived under the assumption $g_{\chi}^V=g_q^V$.
This choice implies that for most of the parameter space
the di-top and di-jet searches for heavy resonances are more sensitive to $g_{\chi/q}^V$ than the mono-jet search for DM.
In \reffig{fig:gqgx} we show how the bounds depend on $g_{q}^V$ and $g_{\chi}^V$ separately,
for values on the mediator and DM masses set at $m_{Z'}=1000\gev$, $\mchi=10\gev$.
The case $g_{\chi}^V=g_q^V$ corresponds to the dashed blue line.
One can see that $Z'$ searches are more constraining than the mono-jet search for a wide choice of couplings,
but the limits switch position when $g_{\chi}^V\gsim 3.5\,g_{q}^V$\, due to the increased branching fraction $\textrm{Br}(Z'\rightarrow\chi\bar{\chi})$.
Therefore, one must always keep in mind that even if the bounds 
presented here are valid over large regions of the parameter space, 
there remains a significant dependence on the 
underlying assumptions.

For $g_{\chi}^V\neq g_{q}^V$ the upper bounds in \reffig{fig:Z_H} 
will move in the ($y_{\chi}$, $\sqrt{g_q^V g_{\chi}^V}$) plane. 
In particular for $g_{\chi}^V < g_{q}^V$ the upper bounds from di-top and di-jet searches will become stronger, constraining smaller values of the product, 
and the monojet upper bounds will become 
weaker. For $g_{\chi}^V>g_{q}^V$ it will be the other way around. 
This can be understood from \reffig{fig:gqgx}: 
if one takes $g_{\chi}^V$ as large as possible, i.e.  $\sim 1.4$, then the upper bound on $g_{q}^V$ derived from the monojet search is $\sim0.15$, as the figure shows. 
One can then evaluate $\sqrt{1.4\cdot 0.15}=0.45$ to get un upper bound on the product. This is essentially the strongest 
the monojet bound can get at the expense of the bounds from the di-jet distribution of the $Z'$. 
The product $\sqrt{g_q^Vg_{\chi}^V}$ is the important quantity for the blind spot and direct detection.

\subsection{Model 2: combining Higgs portal and squarks}

This LSMS features some of the characteristics of SUSY models of DM. In particular 
the bounds can resemble those obtained in cases where the neutralino couples to the SM Higgs and additional heavy Higgs bosons; see, e.g.,\cite{Crivellin:2015bva}. 
An obvious difference with the MSSM is that in the case presented here the DM is a Dirac fermion with free couplings, and the additional scalar is a SM singlet.  

The low-energy Lagrangian of Model~2 is given by the sum of \refeq{higport} and \refeq{squarks}.
There are 6 free parameters,
\be
\left\{ \mchi, m_{\tilde{q}}, m_H, \theta, y_\chi, g_{\tilde{q}}\right\}.
\ee

As before, we fix $\theta=0.2$ because of the mild dependence on $\sin 2\theta$.
We start by fixing the masses of the mediators, $m_{\tilde{q}}=1000\gev$ and $m_H = 600\gev$.
Recall that the latter choice implies that we do not investigate interference effects between the heavy and light Higgs bosons.

The effective amplitudes for DD in this case read
\begin{multline} 
f_n\approx f_p\approx \frac{y_{\chi}\sin 2\theta}{4\,m_{h_{\textrm{SM}}}^2}\left(1-\frac{m_{h_{\textrm{SM}}^2}}{m_H^2}\right)\frac{m_p}{v}\left(\sum_{q=u,d,s}f_{Tq}+\frac{2}{9}\,f_{TG}\right)\\
+\frac{m_p}{m_q}\left(\mathcal{C}_{\textrm{tree}}\sum_{q=u,d,s}f_{Tq}+\mathcal{C}_{\textrm{box}}\,f_{TG}\right)\frac{g_{\tilde{q}}^2}{m_{\tilde{q}}^2-\mchi^2}\,,\label{ampl2}
\end{multline}
where we parametrize the numerical coefficients relative to the tree-level and box-diagram WIMP-quark-squark interactions in the proton 
with $\mathcal{C}_{\textrm{tree}}$ and 
$\mathcal{C}_{\textrm{box}}$, respectively. The numerical value of the sum in parenthesis in the second line of \refeq{ampl2} is 0.019.
Note also that we use the \texttt{micrOMEGAs} default constituent value $m_{q}=0.05\gev$ for the 3 lightest quarks. 

When $y_{\chi}>0$
cancellations in the amplitude for \sigsip\ do not occur. We thus 
here limit ourselves to describing the case $y_\chi<0$, 
so that interference effects create a blind spot for DD searches.
The condition for the blind spot is given by
\be 
y_{\chi}\approx -\left(\frac{2.05\times 10^7\textrm{ GeV}^2}{m_{\tilde{q}}^2-\mchi^2}\right)
\frac{g_{\tilde{q}}^2}{\sin 2\theta\, \left(1-\frac{m_{h_{\textrm{SM}}}^2}{m_H^2}\right)}\,.\label{BS2}
\ee

\begin{figure}[t]
\centering
\subfloat[]{%
\label{fig:a}%
\includegraphics[width=0.45\textwidth]{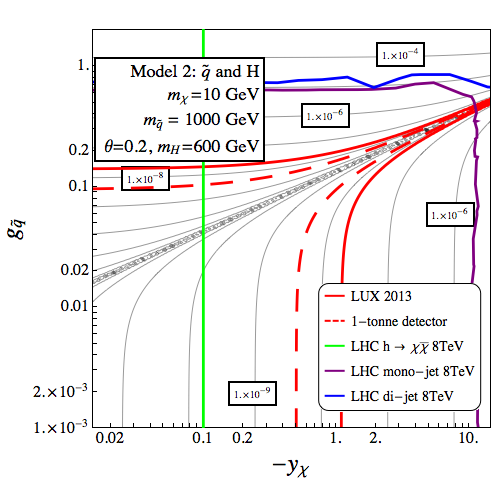}
}%
\hspace{0.07\textwidth}
\subfloat[]{%
\label{fig:b}%
\includegraphics[width=0.45\textwidth]{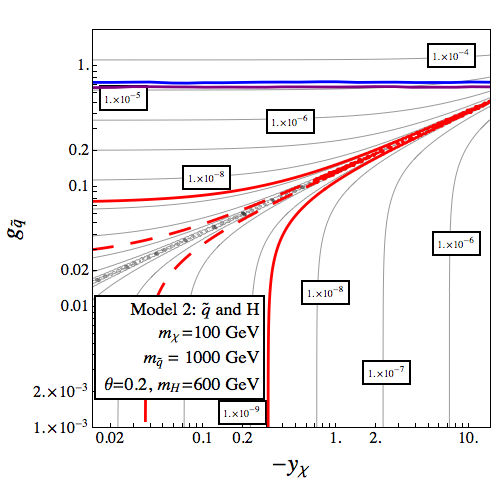}
}%
\caption{(a) Spin-independent  scattering cross section in pb in the ($y_\chi$, $g_{\tilde{q}}$) plane for a model combining 
squark-like mediators and Higgs portal. We set $\mchi = 10\gev$, $m_{\tilde{q}} = 1000\gev$, $\theta = 0.2$, and $m_H=600\gev$. 
The solid red line represents the current 90\%~C.L. upper bound from LUX and the dashed red line is the projected limit 
for tonne-scale detectors in 2017 or so. 
Solid purple line is the 95\%~C.L. upper bound from the 8\tev\ mono-jet search at ATLAS and 
solid blue line gives the 95\%~C.L. upper exclusion bound from direct squark searches with jets and missing energy\cite{Aad:2014wea}. 
Green solid line gives the upper limit from the invisible width of the Higgs boson from the combined CMS/ATLAS analysis.
The full parameter space shown is within reach of 14\tev\ jets+MET and mono-jet searches.
(b) Same as (a) but $\mchi = 100\gev$.}
\label{fig:squark_h_couplings}
\end{figure}

We present in \reffig{fig:squark_h_couplings}(a) contours of \sigsip\ in pb in the 
($y_\chi$, $g_{\tilde{q}}$) plane for the case $\mchi = 10\gev$.
The color code for the bounds is the same as in \reffig{fig:Z_H}.
Additionally, the solid blue line shows the upper bound from the ATLAS 8\tev\ squark search in jets + missing $E_T$\cite{Aad:2014wea} (see also\cite{Khachatryan:2015vra} for the CMS bound).
As was the case for Model~1, the bound on $|y_{\chi}|$ from the invisible width of the Higgs is stronger than the LUX 
bound from DD for this benchmark point.

In \reffig{fig:squark_h_couplings}(b) we show the case with $\mchi=100\gev$, for which the 
bound from the invisible width obviously does not apply. As was the case in \reffig{fig:Z_H}, the bound from mono-jet is less constraining 
for $|y_{\chi}|$ in this case, as the SM Higgs is not produced on-shell.
Thus, the parameter space that is not in reach of underground DD experiments remains essentially unconstrained, 
and this is in particular true for the blind spot, when $\mchi\gsim 62\gev$.

On the other hand, we stress that for the squark mass considered in \reffig{fig:squark_h_couplings}, $m_{\tilde{q}} = 1000\gev$,
the 14\tev\ jets+MET search with 300\invfb\ at ATLAS\cite{ATL-PHYS-PUB-2014-010}, 
as well as the 14\tev\ mono-jet search, are expected 
to exclude the full parameter space shown in the figures. 

\begin{figure}[t]
\centering
\includegraphics[width=0.55\textwidth]{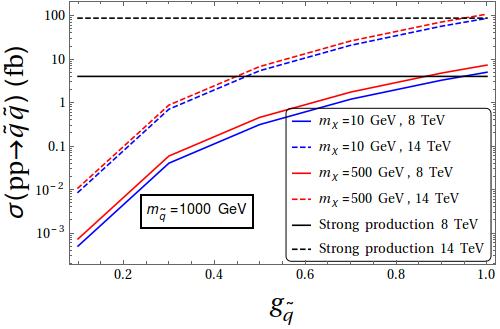}
\caption{Solid blue line shows the cross section for squark production through $t$-channel DM exchange 
at the LHC 8\tev\ for $m_{\tilde{q}}=1000\tev$ and a WIMP mass $\mchi=10\gev$, as a function of the $g_{\tilde{q}}$ coupling.
Solid red line shows the same production cross section when $\mchi=500\gev$. The equivalent cross sections at the LHC 14\tev\ are shown with dashed blue and red line,
respectively. Solid black line shows the cross section for strong squark production at the LHC 8\tev\ and the dashed black line the equivalent strong production cross section at
the LHC 14\tev\cite{LHCSXSECWG}.}
\label{fig:x-sec}
\end{figure}

To show this in detail, we first present in \reffig{fig:x-sec} plots of the squark production cross sections at the LHC 
as a function of the $g_{\tilde{q}}$ coupling for $m_{\tilde{q}}=1000\tev$ and different choices of the WIMP mass. 
For $m_{\tilde{q}}=1000\tev$, the cross sections for strong\cite{LHCSXSECWG} and $t$-channel DM exchange production of the squarks
become of equal size when $g_{\tilde{q}}\approx 0.9$, and this is true at the LHC 8 and 14\tev\ alike. 

\begin{figure}[t]
\centering
\subfloat[]{%
\label{fig:a}%
\includegraphics[width=0.45\textwidth]{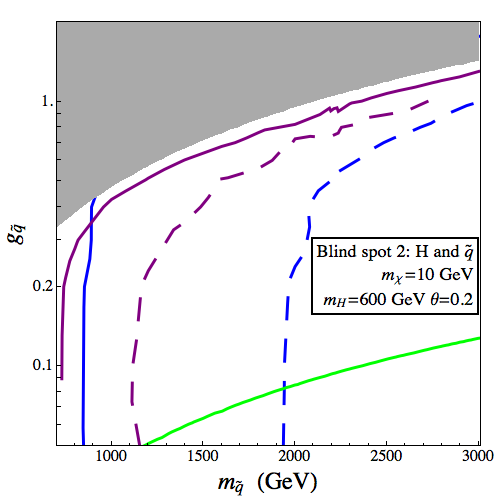}
}%
\hspace{0.07\textwidth}
\subfloat[]{%
\label{fig:c}%
\includegraphics[width=0.45\textwidth]{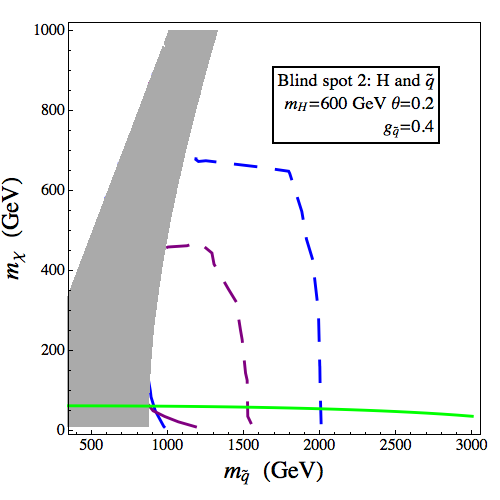}
}%
\caption{(a) Interplay of LHC constraints for the blind spot parametrized by \refeq{BS2} in the ($m_{\tilde{q}}$, $g_{\tilde{q}}$) plane. Here $\mchi=10\gev$, $m_{H}=600\gev$, 
and $\theta=0.2$. 
Solid purple line gives the 95\%~C.L. upper bound from mono-jet at ATLAS, and solid blue line the 95\%~C.L. upper bound from jets + missing $E_T$. 
The corresponding 14\tev\ projections\cite{ATL-PHYS-PUB-2014-007,ATL-PHYS-PUB-2014-010} with 300\invfb\ are given by a dashed purple line and a dashed blue line, respectively. 
The green solid line gives the upper bound from the invisible width of the Higgs in a CMS/ATLAS combined analysis.  
(b) The bounds projected to the ($m_{\tilde{q}}$, \mchi) 
plane for $g_{\tilde{q}}=0.4$ and $m_{H} = 600\gev$. 
Corresponding bounds are now lower limits on the masses.}
\label{fig:squark_h_det}
\end{figure}

Then, in \reffig{fig:squark_h_det}(a) we show the dependence of the bounds on the mediator and DM masses when  
\refeq{BS2} holds. The first feature to note is that, as was the case in the previous model, the invisible branching fraction of the Higgs 
yields the greatest constraint when $\mchi < 62.5\gev$,
with the exception of the region characterized by small $g_{\tilde{q}}$ (how small it should be to evade the bound depends on $m_{\tilde{q}}$).
It is also clear that at 8\tev\ the jets+MET and mono-jet searches dominate in different regions of the parameter space:
when $g_{\tilde{q}}$ is small direct production of the DM is negligible with respect to the strong production of the squarks,
whose cross section is shown as a black solid line in \reffig{fig:x-sec}.
As a consequence the jets+MET search has the greatest sensitivity and excludes a squark mass $m_{\tilde{q}}\lesssim 900\gev$.
As $g_{\tilde{q}}$ increases direct WIMP production becomes more important and the mono-jet bound overtakes the jets+MET bound for $g_{\tilde{q}} \gsim 0.3$.

At 14\tev\ the reach of the mono-jet and jets+MET searches both increase considerably.
Taking the current projections of the backgrounds and cuts\cite{ATL-PHYS-PUB-2014-007,ATL-PHYS-PUB-2014-010} at face value, 
it appears that the jets+MET search will play a more significant role in constraining the parameter space of Model~2 than the mono-jet search for Run~2.

In \reffig{fig:squark_h_det}(b) we show the limits in the ($m_{\tilde{q}}$, \mchi) plane. 
Both the mono-jet and jets+MET limits show the well known characteristic shapes of squark searches; 
however, the region where the squark and WIMP masses are close to degenerate is not present here, as it corresponds
to non-perturbative values of the coupling $y_\chi$\,.
Thus, the limits on squark masses become more robust in the blind spot than in the individual SMS.
Also visible is the apparent weakening of the limit on \mchi\ from $\textrm{Br}(h_{\textrm{SM}}\rightarrow\chi\bar{\chi})$ as the squark mass increases, 
because $y_\chi$ decreases as a function of $m_{\tilde{q}}$.

\subsection{Model 3: combining \boldmath$Z'$ and squarks\label{sec:sqz}} 

In this subsection we consider an LSMS designed to mimic a UV completion 
characterized by an additional $U(1)_X$ symmetry
that remains unbroken down to collider energies (see, e.g.,\cite{Athanasopoulos:2014bba}).
One could imagine that $U(1)_X$ is spontaneously broken by the vev of a scalar field $\Psi$ charged under $U(1)_X$,
giving rise to a light $Z'$ boson which becomes part of the low-energy spectrum. The low-energy spectrum also contains 
scalar particles charged under $SU(3)$ whose origin could be SUSY or something else. The details of the UV completion are 
not important for the phenomenology of the LSMS.

Constructing a gauge invariant combination of the two existing SMS requires some care since there is more than one way 
Model~3 could be constructed without breaking gauge invariance of the full Lagrangian.
A first simple way is to imagine that the DM particle and the quarks have the same charges 
under $U(1)_X$, and the scalar colored particles are instead $U(1)_X$ neutral. 
In this case the Lagrangian is a straight combination of Eqs.~(\ref{veclagr}) and (\ref{squarks}), with a coupling 
$g_{\chi/q}^V\equiv g_{\chi}^V=g_q^V$, and such that the extra scalars do not couple to the $Z'$.
The resulting phenomenological model is described by 5 free parameters: \mchi, $m_{\tilde{q}}$, $m_{Z'}$,  $g_{\chi/q}^V$, and $g_{\tilde{q}}$\,. 
However, this model cannot develop destructive interference between the diagrams with squark exchange and those with 
a $Z'$ mediator, so that for the purposes of this study it is not very interesting.

Another way of constructing a gauge invariant LSMS out of a combination of the SMS with squarks 
and a $Z'$ is the following, which allows the squarks to have the same coupling to the $Z'$ as the quarks, and could be seen as 
an approximation of a full UV theory involving an extended gauge symmetry and a supersymmetric sector.  
One needs two fermion SM singlet DM candidates, $\xi$ and $\zeta$, such that 
$\xi$ is coupled to the $Z'$ like in \refeq{veclagr} and $\zeta$ is coupled to the 
squarks like in \refeq{squarks}. The symmetry is conserved if the fields are charged under $U(1)_X$ according to the following table,
\begin{eqnarray}
\begin{tabular}
[c]{|c|c|c|c|c|c|c|c|}
\hline
 & $\Psi$ & $\xi$ & $\zeta$ & $q_i$ & $\tilde{q}_{i,L/R}$ \\\hline
$U(1)_X$ charge & $+1$ & $+1$ & $0$ & $+1$ & $+1$ \\\hline
\end{tabular}
\end{eqnarray}
where we have normalized all charges to 1 for simplicity, and we leave some freedom in choice of the coupling constants. 
As was the case before we assume that any anomalies are cancelled by additional states which do not effect the phenomenology described here.
The low energy Lagrangian can contain the additional terms,
\be 
\mathcal{L}\supset y_1 \Psi\bar{\xi}\zeta+\frac{1}{2}m_{\xi}\bar{\xi}\xi+\frac{1}{2}m_{\zeta}\bar{\zeta}\zeta+\textrm{ h.c.}\,,
\ee
where $\Psi$ is the field that breaks $U(1)_X$ when it gets a vev $\Psi\rightarrow v_{\Psi}+\psi$, with $\psi$ a decoupled physical scalar. 
After the symmetry is broken, $\xi$ and $\zeta$ mix giving rise to two mass eigenstates: $\chi_1$ and $\chi_2$ which, if 
we assume  $m_{\xi}, m_{\zeta}\ll y_1v_{\Psi}$, are almost mass degenerate with a mass $\mchi=y_1 v_{\Psi}$ and maximal mixing. 

Despite being apparently rather involved, the phenomenological LSMS is characterized by only 6 free parameters,
\be
\left\{ \mchi, m_{\tilde{q}}, m_{Z'},  g_{\chi}^V, g_q^V, g_{\tilde{q}}\right\}\,.
\ee
To reduce this number one can make the additional assumption
$g_{\chi}^V=\pm g_q^V\equiv g_{\chi/q}^V$. 

Destructive interference between different diagrams arises when $g_{\chi}^V=-g_{q}^V$, as
\be
f_n\approx f_p\approx \frac{3}{2}\frac{g_{\chi}^Vg_q^V}{m_{Z'}^2}+\frac{m_p}{m_q}\left(\mathcal{C}_{\textrm{tree}}\sum_{q=u,d,s}f_{Tq}+\mathcal{C}_{\textrm{box}}\,f_{TG}\right)\frac{g_{\tilde{q}}^2}{m_{\tilde{q}}^2-\mchi^2}\,,\label{ampl3}
\ee
see also Eqs.~(\ref{ampl}) and (\ref{ampl2}). 
The condition for the blind spot is thus
\be 
|g_{\tilde{q}}|\approx 2\,\left|g_{\chi/q}^V\right|\frac{\sqrt{m_{\tilde{q}}^2-\mchi^2}}{m_{Z'}}\,.\label{BS3}
\ee 

\begin{figure}[t]
\centering
\subfloat[]{%
\label{fig:a}%
\includegraphics[width=0.45\textwidth]{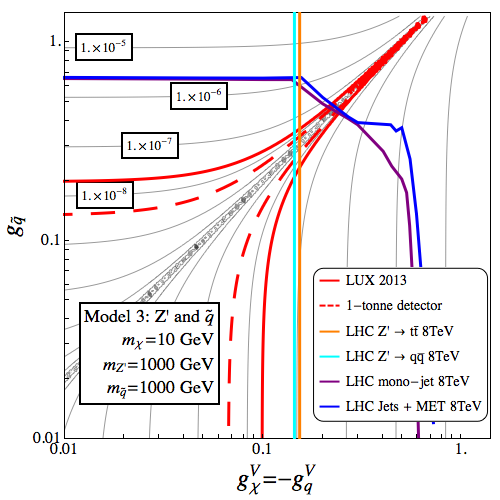}
}%
\hspace{0.07\textwidth}
\subfloat[]{%
\label{fig:b}%
\includegraphics[width=0.45\textwidth]{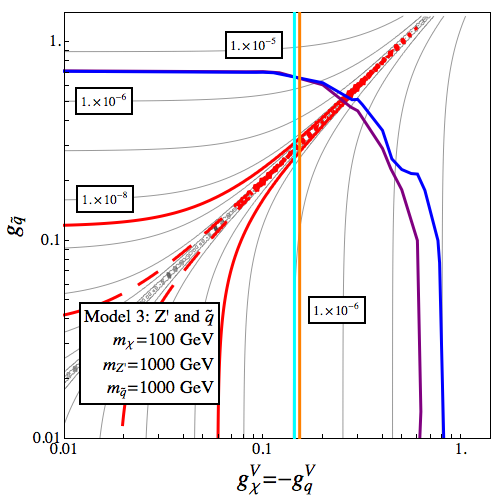}
}%
\caption{(a) Spin-independent scattering cross section in pb in the 
($g_{\chi}^V=-g_{q}^V$, $g_{\tilde{q}}$) plane for a combined $Z'+\textrm{squark}$ mediator simplified model. 
The masses are fixed at $\mchi = 10\gev$, $m_{Z'}=1000\gev$, and $m_{\tilde{q}}=1000\gev$. 
The solid red line shows the current upper limit from LUX and the 
dashed red line the projected limit for tonne-scale underground detectors. 
Solid purple line is the upper limit from the 8\tev\ mono-jet search at ATLAS and
solid blue line is the upper limit from the 8\tev\ jets+MET search at ATLAS.
Solid orange line is the ATLAS limit on $Z'$ resonances from the invariant $t\bar{t}$ mass distribution at 8\tev, while the solid cyan line 
gives the equivalent bound from the $\bar{q}q$ distribution.
The full parameter space shown is within reach of 14\tev\ jets+MET and mono-jet searches.
(b) Same as (a) but $\mchi=100\gev$.}
\label{fig:Z_squark}
\end{figure}

We plot in \reffig{fig:Z_squark}(a) contours of \sigsip\ in pb in the 
($g_{\chi/q}^V$, $g_{\tilde{q}}$) plane for $\mchi = 10\gev$. The masses of the mediators are here set at $m_{\tilde{q}}=1000\gev$
and $m_{Z'}=1000\gev$. The color code is the same as in the previous figures.
The case with $\mchi = 100\gev$ is shown in \reffig{fig:Z_squark}(b). One can see again that the collider bounds
barely move by changing the DM mass, but DD bounds reach their close-to-maximal strength 
when $\mchi = 100\gev$.

Note that the mono-jet and jets+MET ATLAS searches yield very comparable bounds for this choice of mediator masses.
Note also that, for these mediator masses, the 14\tev\ projected reach with 300\invfb\ for both searches
covers the full parameter space, so that the limit does not appear in the figure.   

\begin{figure}[t]
\centering
\subfloat[]{%
\label{fig:a}%
\includegraphics[width=0.45\textwidth]{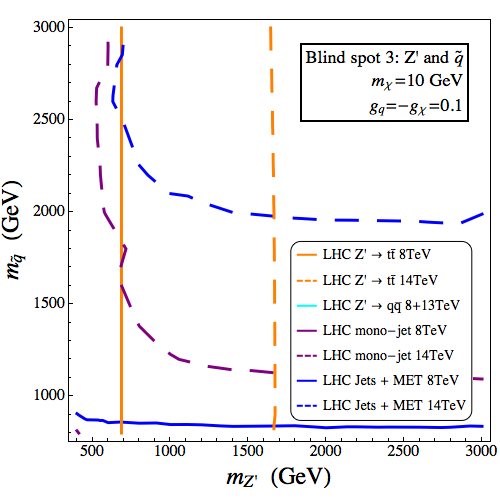}
}%
\hspace{0.07\textwidth}
\subfloat[]{%
\label{fig:b}%
\includegraphics[width=0.45\textwidth]{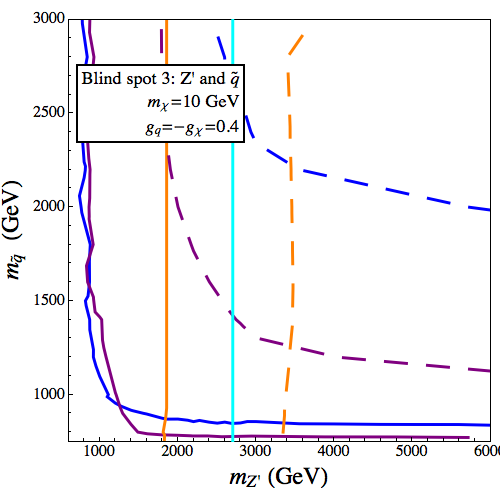}
}%
\caption{(a) Interplay of LHC constraints for the blind spot parametrized by \refeq{BS3} in the ($m_{Z'}$, $m_{\tilde{q}}$) plane. Here $\mchi=10\gev$, $g_{q}^V=-g_{\chi}^V=0.1$. 
Solid purple line gives the 95\%~C.L. lower bound from mono-jet at ATLAS, and solid blue line the 95\%~C.L. lower bound from jets + missing $E_T$. The corresponding 14\tev\ projections with 300\invfb\ are given by a dashed purple line and a dashed blue line, respectively. The solid orange line gives the current 95\%~C.L. lower bound on heavy resonances from the di-top distribution. The corresponding projected reach at LHC 14\tev\ with 300\invfb\ is shown with a dashed orange line. The solid cyan line gives the bound on $Z'$ from the $\bar{q}q$ invariant mass distribution at the LHC 13\tev, 4\invfb. (b) Same as (a) but $g_{q}^V=-g_{\chi}^V=0.4$.}
\label{fig:Z_squark_det}
\end{figure}

The dependence of the collider bounds on the mediators' mass when we confine ourselves to the blind spot 
for DD experiments (i.e. \refeq{BS3} holds) is shown in \reffig{fig:Z_squark_det}. In \reffig{fig:Z_squark_det}(a) we show 
the interplay of the collider bounds on the ($m_{Z'}$, $m_{\tilde{q}}$) plane for $g_{\chi/q}^V=0.1$.
The figure shows that a squark mass below $\sim 900\gev$ is excluded by the jets + missing energy ATLAS search, independently
of the couplings, because squarks are produced via strong interactions.
On the other hand a $Z'$ mass below $\sim 700\gev$ is excluded by the di-top search for heavy resonance when $g_{\chi/q}^V=0.1$.
The line moves to the right for larger couplings, as is shown in \reffig{fig:Z_squark_det}(b), where we present the case with 
$g_{\chi/q}^V=0.4$.

The reach of the mono-jet and jets+MET searches at the LHC 14\tev, shown as a dashed purple and dashed blue line, respectively, 
are very sensitive to the size of $g_{\chi/q}^V$\,. In general, our projections, based on\cite{ATL-PHYS-PUB-2014-007,
ATL-PHYS-PUB-2014-010}, show a significantly greater 
reach for the jets+MET search, which for $g_{\chi/q}^V=0.4$ can probe the squark mediator mass up to $\sim2000\gev$. 

Conversely, the reach of the di-top search for heavy resonances will allow one to probe the $Z'$ mass up to more than 3000\gev\
for this coupling, showing again, that a complementary use of different detection strategies can constrain a large part of the parameter space invisible in DD experiments.

\section{\label{sec:summary} Summary and conclusions}

In this paper we have considered LSMS as simple extensions to dark matter simplified models,
with the goal of mimicking some of the characteristics of more realistic models without introducing an excessively large number of parameters. 
Our starting assumption is that, to some extent, many of the features of complex models that cannot 
be explored in a single SMS framework, like interference of different diagrams that result in suppressions of the DM scattering
cross section, or complementary use of different experimental strategies to explore a variegated parameter space from different angles, 
can instead be employed in LSMS built from simple combinations of existing SMS.

We thus considered three cases characterized by a Dirac fermion WIMP coupled to more than one mediator: 
1) heavy vector mediator and Higgs portal; 2) squark-like mediator and Higgs portal; and 3) 
squark-like mediator and heavy vector mediator.
We have limited ourselves to the analysis of SMS characterized by a sizeable spin-independent scattering cross section, \sigsip,
and to a range of DM masses for which a direct comparison of the limits from the LHC with the limits from DD in underground 
laboratories is possible, i.e., we have not considered cases with $\mchi<10\gev$. 
In each case we have confronted the models with the present bounds from DD at LUX, 
and a number of constraints from Run~1 at the LHC: mono-jet searches, direct searches for heavy resonances in the di-top and di-jet channels, 
invisible width of the SM Higgs boson, and searches for colored 
heavy particles in jets with missing $E_T$\,. 

In particular, we dedicated special attention to the parameter space showing interference 
between different diagrams, which gives rise to blind spots for DD experiments.
For these blind spots we have investigated to what extent interplay between collider limits on different particles and couplings
can be used for a full exploration of the parameter space and we have additionally considered projections for the LHC 14\tev\ run. 

As is well known, in general for DM mass ranges such that $\mchi\geq10\gev$, 
the LUX upper bound on \sigsip\ constrains the coupling constants of WIMP SMS 
by at least one order of magnitude more strongly than any of the LHC searches considered here. 
Even more strikingly, we find that the present bound from LUX also outperforms projected bounds for 300\invfb\ at the LHC 14\tev\
in mono-jet searches and the projected reach of searches for heavy vector resonances in most cases. 
 
A few exceptions to this rule can however be found. 
One is that in models with a Higgs portal, for $\mchi\lesssim 1/2\,m_{h_{\textrm{SM}}}$ the coupling between the Higgs bosons 
and the WIMP are also strongly constrained by the invisible width of the SM Higgs boson, in agreement with what was observed in\cite{Baek:2014jga}.
The other exception, which comprises the central part of this work, 
is that in regions of the parameter space where \sigsip\ is suppressed by interference effects  
a combination of LHC searches can effectively place strong limits,
especially at the end of Run~2.  

In particular, we found that the following features apply to detections of a blind spot: 
\begin{itemize}
\item The model involving a $Z'$ and Higgs portal (Model 1) is at present not constrained at all by mono-jet searches in the blind spot
if $g_{\chi}^V=g_q^V$. Moreover, under this assumption, the projections for the LHC 14\tev\ show that searches for heavy $Z'$ resonances will 
constitute the most effective strategy, among the ones considered here, to probe this part of the parameter space.
Mono-jet searches, even with 300\invfb\ in the 14\tev\ run, may be competitive with searches for $Z'$
only for cases with $g_{\chi}^V\gg g_q^V$\,.    
\item In the two models (Models 2 and 3) involving squark-like mediators the bounds from mono-jet and jets+MET searches on the coupling 
$g_{\tilde{q}}$ are at present comparable. However, according to the collaborations' simulations\cite{ATL-PHYS-PUB-2014-007,ATL-PHYS-PUB-2014-010}, 
the constraining power of jets+MET searches at the 14\tev\ LHC for the blind spots significantly outperforms the expectations for mono-jet searches. 
\end{itemize} 

In general we find that the complementarity of different search strategies is crucial to obtaining the best constraints on DM at the LHC. 
Many motivated models are not necessarily constrained by direct detection experiments and thus demand careful attention at colliders.
In this paper we have shown some first developments towards describing such scenarios in terms of a less-simplified model framework that can be investigated at the LHC.

\bigskip
\noindent \textbf{Acknowledgments}
\medskip

\noindent We would like to thank Arindam Chatterjee for helpful comments on the manuscript. 
  A.C. would like to thank Tanumoy Mandal for some helpful discussions regarding \texttt{DELPHES} and 
  the BayesFITS Group at the National Centre for Nuclear Research for hospitality and support in the early stages of this work.
  This work has been funded in part by the Welcome Programme
  of the Foundation for Polish Science. The work of A.C. is partially supported by funding available 
  from the Department of Atomic Energy, Government of India, for the Regional Centre for Accelerator-based Particle Physics 
  (RECAPP), Harish-Chandra Research Institute. K.K. is supported in part by the EU and MSHE Grant
  No. POIG.02.03.00-00-013/09.
  L.R. is also supported in part by a Lancaster-Manchester-Sheffield Consortium for Fundamental Physics under STFC grant ST/L000520/1.
  The use of the CIS computer cluster at the National Centre for Nuclear Research is gratefully acknowledged. 

\bigskip

\bibliographystyle{JHEP}

\bibliography{BF_12}

\end{document}